\documentclass[12pt]{spieman} 

\usepackage[super]{nth}
\usepackage{savesym}
\usepackage[T1]{fontenc}
\savesymbol{tablenum}
\usepackage{inputenc}
\usepackage[range-units=brackets,product-units=single,multi-part-units=single,range-phrase={-},separate-uncertainty,print-unity-mantissa=false]{siunitx}
\usepackage{multirow}
\usepackage{longtable}
\usepackage{amsmath,amsfonts,amssymb}
\usepackage{graphicx}
\usepackage{setspace}
\usepackage{tocloft}
\usepackage{lineno}

\usepackage[dvipsnames]{xcolor}

\graphicspath{{./}{figures/}}

\title{Using the XMM-Newton Small Window Mode to investigate systematic uncertainties in the particle background of X-ray CCD detectors}

\author[a,*]{Gerrit Schellenberger}
\author[a]{Ralph Kraft}
\author[a,b]{Paul Nulsen}
\author[c]{Eric D. Miller}
\author[c]{Marshall W. Bautz}
\author[c]{Catherine E. Grant}
\author[d]{Dan Wilkins}
\author[d,e,f]{Steven Allen}
\author[g]{Silvano Molendi}
\author[h]{David N. Burrows}
\author[h]{Abraham D. Falcone}
\author[i]{Valentina Fioretti}
\author[c]{Richard F. Foster}
\author[j]{David Hall}
\author[j]{Michael W. J. Hubbard}
\author[k]{Emanuele Perinati}
\author[d]{Artem Poliszczuk}
\author[l]{Arne Rau}
\author[c]{Arnab Sarkar}
\author[c]{Benjamin Schneider}

\affil[a]{Center for Astrophysics $|$ Harvard \& Smithsonian, 60 Garden St., Cambridge, MA 02138, USA}
\affil[b]{ICRAR, University of Western Australia, 35 Stirling Hwy, Crawley, WA 6009, Australia}
\affil[c]{MIT Kavli Institute for Astrophysics and Space Research, 77 Massachusetts Avenue, Cambridge,
MA 02139, USA}
\affil[d]{Kavli Institute for Particle Astrophysics and Cosmology, Stanford University, 452 Lomita Mall,
Stanford, CA 94305, USA}
\affil[e]{Department of Physics, Stanford University, 382 Via Pueblo Mall, Stanford, CA 94305, USA}
\affil[f]{SLAC National Accelerator Laboratory, 2575 Sand Hill Road, Menlo Park, CA 94025, USA}
\affil[g]{INAF - IASF Milano, Via A. Corti 12, 20133 Milano, Italy}
\affil[h]{Pennsylvania State University, Department of Astronomy and Astrophysics, University Park, Pennsylvania, United States}
\affil[i]{INAF Osservatorio di Astrofisica e Scienza dello Spazio di Bologna, Bologna, Italy}
\affil[j]{The Open University, Centre for Electronic Imaging, Milton Keynes, United Kingdom}
\affil[k]{Institut für Astronomie und Astrophysik, Eberhard Karls Universität 
Tübingen, 72076 Tübingen, Germany}
\affil[l]{Max Planck Institute for Extraterrestrial Physics, Giessenbachstrasse 1, 85748 Garching, Germany}

\cftpagenumbersoff{figure}
\cftpagenumbersoff{table} 
\begin{document} 
\maketitle

\begin{abstract}
The level and uncertainty of the particle induced background in CCD detectors plays a crucial role for future X-ray instruments, such as the Wide Field Imager (WFI) onboard Athena. To mitigate the background systematic uncertainties, which will limit the Athena science goals, we aim to understand the relationship between the energetic charged particles interacting in the detector and satellite, and the instrumental science background to an unprecedented level. In addition, we characterize the temporal variability of the instrumental background from minutes to years. 
These particles produce easily identified ``cosmic-ray tracks'' along with less easily identified signals produced by secondary particles, e.g., X-rays generated by particle interactions with the instrument and indistinguishable from genuine sky X-rays. 
We utilize the Small Window Mode of the PN camera onboard XMM-Newton to understand the time, spatial and energy dependence of the various background components, particularly the particle induced background. 
While the distribution of particle events follows expected detector readout patterns, we find a particle track length distribution inconsistent with the simple, isotropic model. We also find that the detector mode-specific readout results in a shifted Cu fluorescent line. 
We illustrate that on long timescales the variability of the particle background correlates well with the solar cycle. This 20-year lightcurve, can be reproduced by a particle detector onboard Chandra, the HRC anti-coincidence shield. We conclude that the self-anti-coincidence method of removing X-ray-like events near detected particle tracks in the same frame can be optimized with the inclusion of additional information, such as the energy of the X-ray. The results presented here are relevant for any future pixelated X-ray imaging detector, and could allow the WFI and similar instruments to probe to truly faint X-ray surface brightness.

\end{abstract}


{\noindent \footnotesize\textbf{*}\linkable{gerrit.schellenberger@cfa.harvard.edu} }

\begin{spacing}{1}   

\section{Introduction} \label{ch:intro} 
Many science goals carried by instruments of future X-ray observatories, such as the Wide Field Imager (WFI)\cite{Nandra2013-dg,Rau2013-qp,Meidinger2017-kg} onboard Athena, are driven by the detection and modeling of faint, diffuse sources, such as galaxy cluster outskirts. While measurements with current X-ray observatories are often limited by photon statistics, observatories with large collecting areas will mostly be impacted by systematic uncertainties. 
As shown, e.g., by \citenum{Leccardi2008-tp}, the systematics of the background, especially the non-X-ray background (NXB), limit the feasibility of measurements on cluster outskirts, independent of the statistical significance of the measurement: Once the signal is below the background, longer observing times cannot make up for the level of uncertainty in the background modeling. This is not unique to cluster outskirts, but applies to any faint source, such as, distant black holes, or detecting the first galaxy clusters and groups, both major science drivers of Athena/WFI\cite{Ettori2013-nm,Georgakakis2013-ry,Nandra2013-dg}. 
Therefore, the level and properties of the background produced by high-energy galactic cosmic ray (GCR) particles plays a crucial role in the design of future X-ray observatories. 

We classify the NXB as any feature read out by the detector, that is unrelated to sky X-rays focused by the telescope, such as GCRs\cite{Neher1962-wf}. We can broadly divide the particle interactions into primary interactions of the (mostly) GeV protons with the detector, and events from secondary particles. 
The former leave ``tracks'' in the detector that are easily identifiable, and are typically removed during onboard processing, while the latter are electrons or fluorescent X-rays caused by GCRs interacting with the housing and other material surrounding the detector. 
Soft protons emitted by the sun would also fall under the NXB definition, but are unrelated to GCRs and can be filtered out, e.g., with a magnetic diverter\cite{Ferreira2018-cw,Fioretti2018-gj,Breuer2022-bj, Galgoczi2022-vq}, or a thick layer of aluminum (closed filter position, \citenum{Katayama2004-kg}). 

While the primary GCR interactions deposit a charge signature that is very characteristic, both in terms of amplitude and pattern distribution, the events from secondaries, however, can mimic the energy and pattern of sky X-rays and, so, constitute the major component of NXB in X-ray observations. 
This NXB typically starts to dominate over other background components (e.g., the Galactic foreground emission, and X-ray emission of unresolved point sources) above 2\,keV. Any mitigation of the NXB through simple background subtraction or modeling is problematic since the NXB is variable spatially, in energy and in time. 
Additionally, detectors such as the XMM-Newton PN also have a significant noise contribution from electronic readout artifacts\cite{Dennerl2004-fa} below $\sim \SI{300}{eV}$.

As part of a larger effort within the Athena WFI Background working group (see also \citenum{Grant2018-ld,Bulbul2020-zf,Grant2020-vu,Wilkins2020-xc,Miller2022-yy,Sarkar2024-rx}) we aim to characterize the NXB by utilizing the Small Window Mode (SWM) of the European Photon Imaging Camera PN detector onboard XMM-Newton. This enables us to understand the time, spatial and energy dependence of the particle induced background component, which is critical to verify cosmic ray simulations with Geant4. Its similarity to the Athena/WFI in terms of pixels size, a short frame time, and the availability of all (normally rejected background) events, makes the XMM-Newton PN SWM a unique tool for these studies. While typically used for observations of bright point sources, such as quasars, XMM-Newton archive also hosts a wealth of data taken in SWM between science observations, when the satellite was slewing to the next target. These several hundred exposures, each a few kilseconds long, are ideal for background investigations due to the lack of a bright science target near the center of the field. 
While in a previous study, \citenum{Bulbul2020-zf} investigated a similar dataset in the limited energy band from $\SIrange{2}{7}{keV}$, a lot more information can be extracted by, a) utilizing the full energy band, b) including more observations in various filter configurations, and c) comparing these data with data from other instruments, such as the Chandra High Resolution Camera (HRC) anti-coincidence shield rates.

Our goal is to reach a better understanding of the GCR induced background in X-ray CCD observations to both, lower the level of this particle background, and limit the systematic uncertainties connected with it. 
The latter is required so techniques that account for the background in X-ray observations become more reliable (e.g., background subtraction), due to its variable nature. 
PN allows observations in SWM,  that operates a small area of the detector (about 3\% of the total area), but does not apply onboard minimum ionizing particle (MIP) thresholding, i.e., all events processed are retained in the final event file. 
Self anti-coincidence (SAC) is an important tool to mitigate the particle induced background, and therefore illustrate and optimize the methodology. {\citenum{Miller2022-yy}} demonstrated through detailed Geant4 simulations the prospects of Geant4 for the Athena WFI. 
While many of our results feed into the application of SAC, it is not the only focus of this study. We want to emphasize that only a thorough understanding of all manifestations of the particle background allows us to develop algorithms that reduce the systematic uncertainties in X-ray observations. SAC is one way of directly lowering the background, but the knowledge of the time variability of particles for example, greatly helps to develop or adjust algorithms that rely on the precise particle background level.

In Section \ref{ch:methods} we describe our XMM-Newton and Chandra HRC data reduction and analysis procedure. Section \ref{ch:pbk} explores the spatial, spectral and temporal characteristics of the NXB components, and compares the time variability of primaries and secondaries with the Chandra HRC shield. 
Section \ref{ch:sac} links the previous results with an optimized self-anti coincidence filtering method, which enables a significant reduction of the background level, especially for detectors like the Athena/WFI. We summarize our findings in Section \ref{ch:summary}. 

\section{Data and Analysis} \label{ch:methods} 
\subsection{XMM-Newton PN}
\subsubsection{Data selection}
The PN SWM only reads an array of 64x63 pixels on CCD 4 of PN, and is our operating mode of choice to trace particle interactions with the detector. It also operates with a short frame time of $\SI{5.6718}{ms}$, and the PN has a pixel size of $\SI{150}{\mu m}$\cite{Briel2003-co}. Both values are similar to those expected for the Athena WFI\cite{Rau2013-qp}. 
A large number of relatively short (typically few ks) PN SWM observations were taken during slews between science target observations. In many cases the filter wheel was in the closed position (FWC) for these slew observations in SWM, making them ideal to study only particle interactions with the detector, with no celestial X-rays hitting the detector. 

The archival PN SWM slew observations start in the year 2002, and span over 20 years. On average 97\,ks per year of data are taken in SWM slews, of which 80\% are the FWC observations that are particularly useful for our study.   
We process a total of 665 SWM observations, including the 308 analyzed in \citenum{Bulbul2020-zf}. 

Lastly, slew observations in full frame mode (FF) include none of the detailed information on particle interactions kept for SWM, but they  can be combined with the knowledge gained from SWM on the particle background to provide a better understanding of soft proton flares (e.g., \citenum{Walsh2014-mb}, and section \ref{ch:softprotons}). 
We analyzed 2666 FF observations from slews, which add to a total livetime of 7.4\,Ms. This dataset is unique as the slew mode provides an average sky exposure for every observation, and even slews across bright sources in the sky  will not bias the overall measurement. 

\subsubsection{Data processing}

\begin{figure}
    \centering
    \includegraphics[trim=750pt 18pt 720pt 170pt,clip,width=0.49\textwidth]{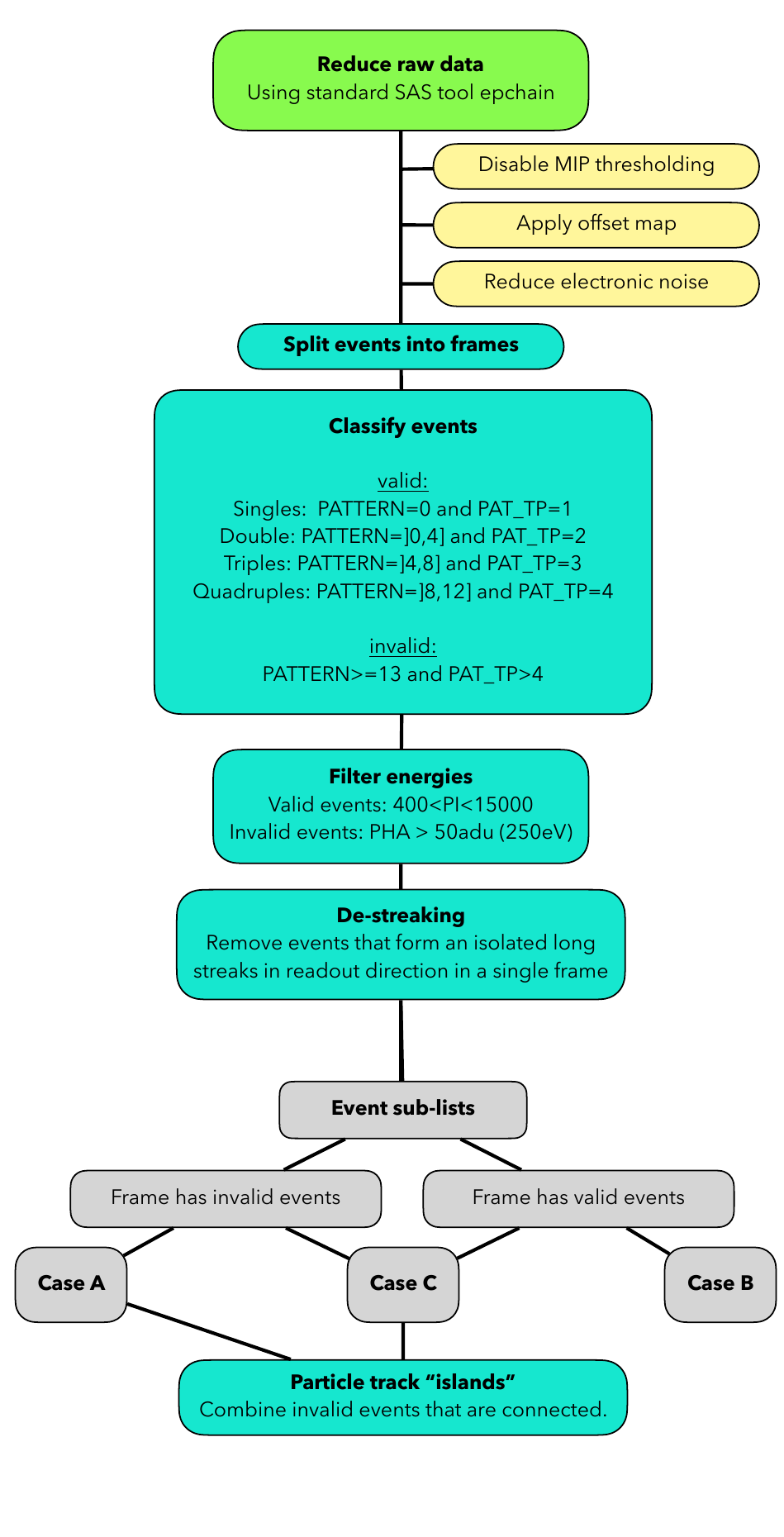}
    \caption{Illustration of the processing of the XMM-Newton PN Small Window Mode Slew datasets to separate particle-track-only frames (Case A), Sky X-ray events (Case B), and mixed frames (Case C).}
    \label{fig:flow}
\end{figure}
We process the raw XMM-Newton PN data using the Science Analysis Software (SAS\cite{Gabriel2004-ty}, version 21), and incorporate some non-standard parameters and tasks. We illustrate our processing pipeline in Fig. \ref{fig:flow}. 
First, we download the Observation/Slew Data Files (ODF/SDF) from the XMM-Newton archive server, and process it using the Current Calibration Files (CCF) from 2023-11-01. 
We follow the default steps using the standard SAS tasks, including \verb|cifbuild|, and \verb|odfingest|. To create the event list with \verb|epchain|, we apply the parameter \verb|anmip=4095|, to disable the MIP thresholding and include all particle events in the output event file. 
Offset maps are typically used to subtract the energy offset for each pixel. However, slew observations do not take offset maps before the exposure, and we make use of the \href{https://xmm-tools.cosmos.esa.int/external/sas/current/doc/epreject}{runepreject} algorithm\cite{Dennerl2004-fa} to resolve this issue. 
We also apply the \verb|epnoise| algorithm\cite{Dennerl2004-fa}, which removes frames that are dominated by electronic noise below 150\,eV. 

To not introduce any biases in our analysis, we turned off the randomization and badpixel detection algorithms. The processed fits file contains an extension  with the resulting event list(HDU 1), and a further extension containing a list of all the frames with corresponding timestamps (HDU 4). 
We found a total of 665 SWM slew observations in the archive made up to the end of November 2023, 519 with the filter in the closed position, 139 with the medium optical blocking filter, and 7 with the FWC and the onboard calibration source (calclosed). 
In the following we exclude observations with a very high radiation environment: The median Advanced Composition Explorer (ACE) EPAM proton flux (112--187\,keV) has to be below $\SI{2450}{proton\,s^{-1}\,cm^{-2}\,sr^{-1}\,MeV^{-1}}$. 
This excludes 32 FWC and 7 medium filter observations taken during time periods with high solar activity. 

\begin{figure}
    \centering
    \includegraphics[width=0.99\textwidth]{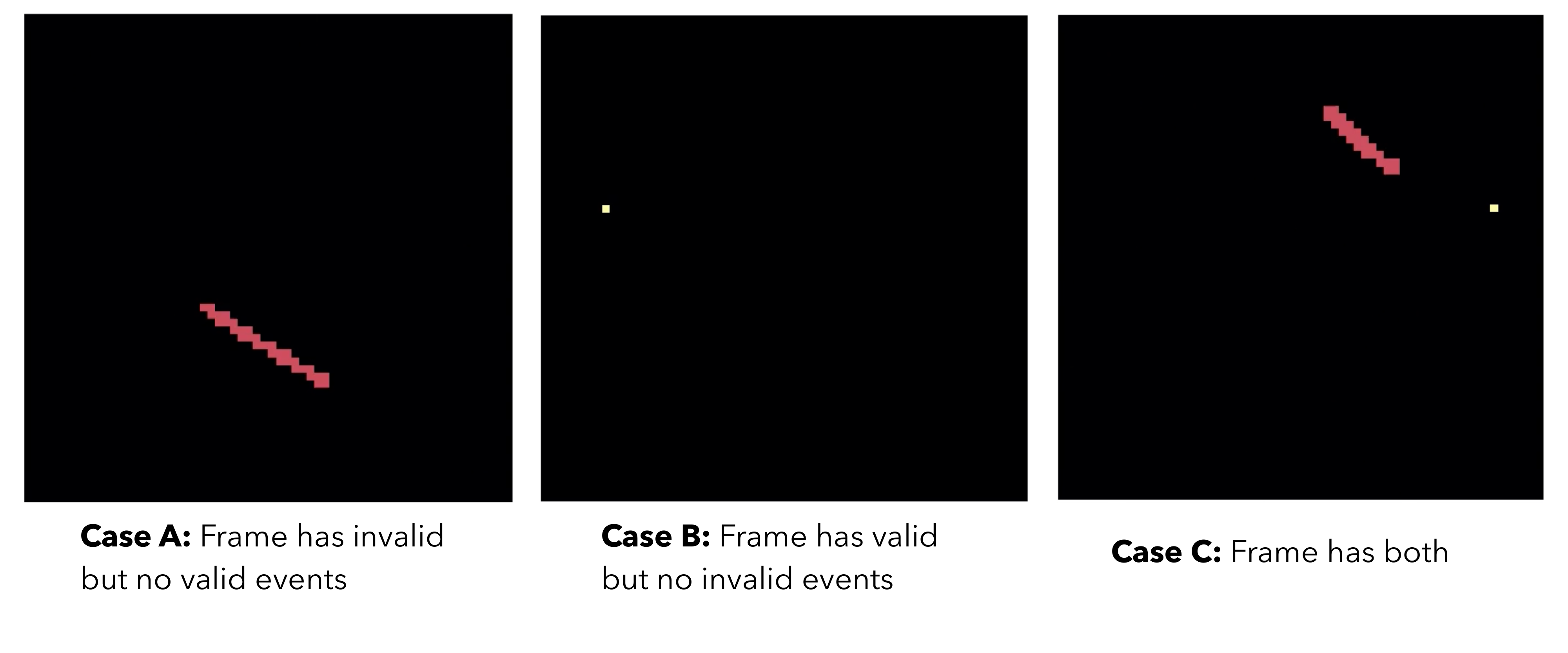}
    \caption{Examples for the different frame types in the PN/SWM data that contain either valid events (yellow), or invalid events (red). The black box marks the active pixel region. }
    \label{fig:frames}
\end{figure}
The remaining 626 slew observations consist of a total of $\num{355495332}$ frames, summing to more than 2\,Ms of observing time, with $\num{5028119}$ frames ($\SI{1.4}{\%}$) containing events. We list all observations in Tab. \ref{tab:SWM}. 
We then classify the events as valid or invalid, based on the energy and pattern: valid events have either a single, double, triple, or quadruple charge pattern distribution based on the \href{https://xmm-tools.cosmos.esa.int/external/sas/current/doc/emevents/node4.html}{SAS classification}. 
Any event with a pattern larger than quadruple (\verb|PAT_TYP|$>4$ and \verb|PATTERN|$\geq 13$) is classified as invalid. 
We further filter the valid and invalid events in energy or pulse height amplitude (PHA), where valid events are required to have an energy (PI column) between 400 and $\SI{15000}{eV}$, and the invalid events need to have a PHA of at least $\SI{50}{adu}$, equal to approximately $\SI{250}{eV}$ (see Fig. \ref{fig:flow} for an overview of the processing). However, each frame with invalid events must at least contain one event with energy of more than $\SI{16}{keV}$, indicating a MIP. Otherwise the frame is discarded. 
We noticed that in some frames a number of invalid events are arranged in a line following the readout direction. This happens more often than expected for the rare case of perfectly aligned particle tracks, and these detections are likely due to artifacts from the readout. We remove these events from our final event list whenever the length of the streak is at least 5 pixels in the readout direction, and no events are detected in adjacent columns. 
We note that since readout streaks can appear in consecutive frames, we apply the method above to the combination of two consecutive frames, which has twice the number of rows of a single frame. 
In total we find that $\num{3935200}$ frames have only invalid events (Case A frames, $\SI{1.1}{\%}$), $\num{1045910}$ frames have only valid events (Case B frames, $\SI{0.3}{\%}$), and $\num{47009}$ frames have both, valid and invalid events (Case C frames, $\SI{0.01}{\%}$). 
The vast majority (99.94\%) of all Case A frames contain more than two or more invalid events. 
We illustrate the different frame types in Fig. \ref{fig:frames}. 
Invalid events are mostly caused by GCRs passing through the detector and depositing charge in connected pixels. These particle tracks can be further classified using a simple segmentation algorithm that outputs the energy weighted centroid, the total energy and number of pixels in the particle track.
We confirmed that the majority ($>97\%$) of the particle tracks are well approximated by a rectangular box, with two or less pixels deviations.

\subsection{Chandra High Resolution Camera}
HRC is one of the two focal plane instruments onboard Chandra\cite{Murray2000-lc}. 
The HRC consists of microchannel plates to detect X-rays, and is surrounded on 5 sides by a plastic scintillator, with photomultiplier tubes (PMTs) as detectors, that serves as an anti-coincidence shield to reduce the background from energetic particles\cite{ODell2005-fz}. 
In order to detect high levels of solar particle fluxes that could be harmful for the instruments onboard Chandra, the HRC shield also records data when the HRC instrument is not in the focus. 
However, since late 2020 the shield is no longer operated when the HRC is not actively observing. 
The recorded shield rates are several 1000 counts per second, recorded at 32.8\,s time intervals. This implies a negligible statistical uncertainty at minute-length time bins, while it has continuous records of the particle environment since the beginning of the Chandra mission (except for radiation zone passages lasting about 12 hours, and a few spacecraft safing events, losing about 17\% of time) until 2020. 
The HRC shield count rates constitute one of the most valuable datasets to analyze the evolution and correlation of particles that induce the non-X-ray background. 

We retrieved the data from the Chandra X-ray Center, and applied a median filter to clip very high values that are due to corruption of the secondary science data. 

\section{Characterizing the nature of the particle background}\label{ch:pbk}
Systematic uncertainties arising from the unpredictability of the NXB in X-ray observations can bias, and therefore limit, the science goals{\cite{Eraerds2021-je}}. 
While energetic particles that interact with the detector produce the easily identifiable ``cosmic-ray tracks'', secondary X-ray-like events are difficult to identify and remove. 
Such secondaries often produce identical signals to the X-rays focused by the optics and traditional filtering algorithms cannot reliably remove them. 
We develop and optimize algorithms and methods to remove these secondary X-ray-like events, which is strongly dependent on the accurate knowledge of the temporal, spatial, and spectral variation and distribution of both, the signatures of primary interactions (particle tracks) as well as the X-ray-like events from secondaries. 

\subsection{Spatial distribution of particle tracks in the FWC slew data}\label{ch:spatialvar}
\begin{figure*}
    \centering
    \includegraphics[width=0.87\textwidth]{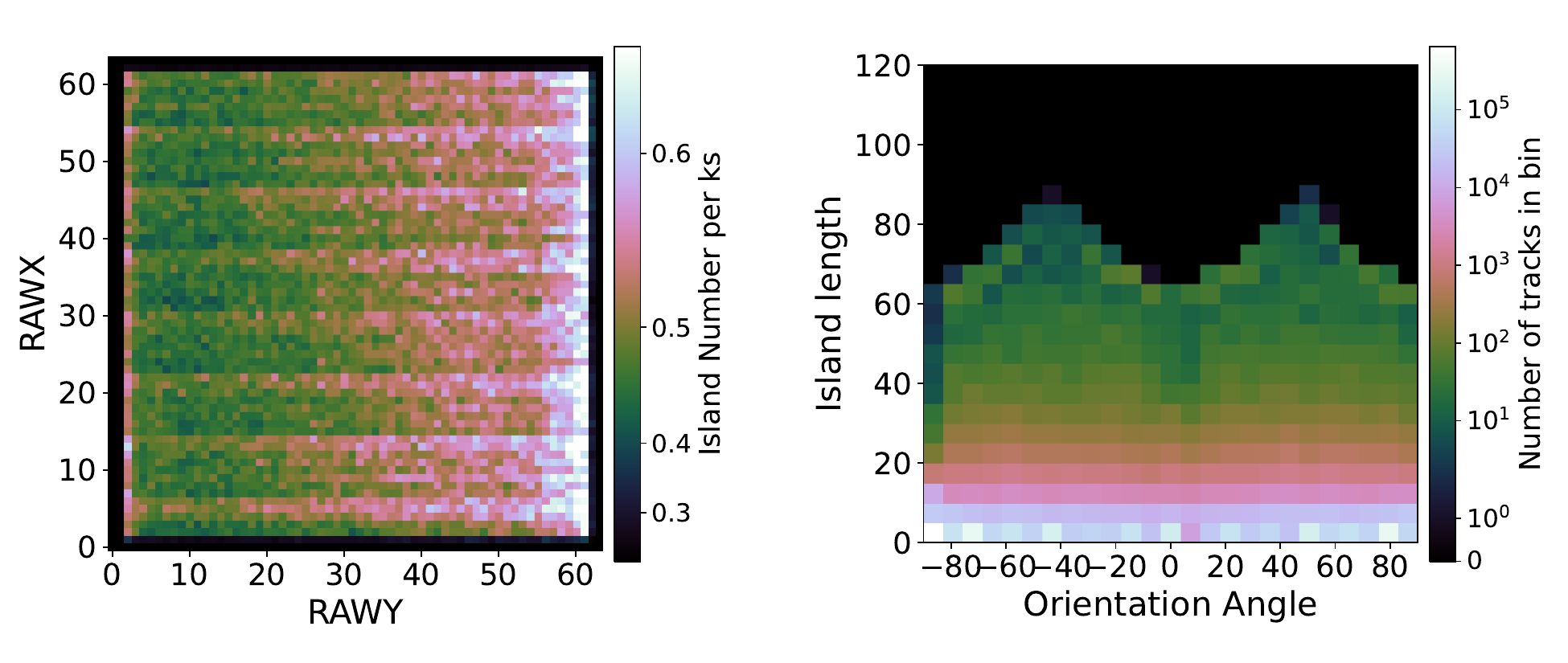}
    \caption{Left: Average number of the particle tracks (islands) as a function of their centroid distribution. The colorscale is in units of island centers in each pixel per ks. Right: Histogram of the island orientation and length distribution. The colorscale shows the number of islands in each bin.}
    \label{fig:island_dist}
\end{figure*}
In order to understand the distribution of particle tracks, which consist of invalid events, we define event islands, where all invalid events are merged into a single shape as long as they are not separated by more than $\sqrt{2}$ pixel. These islands are characterized by a centroid, a major and minor axis, and the rotation angle (orientation). Therefore, a single frame can have more than one island, if the closest separation is more than $\sqrt{2}$ pixel. 
We employ a simple segmentation algorithm{\cite[Chapter~2]{Jain1995-xz}} that provides these island parameters. 
We investigate the distribution properties in the following. 

\begin{figure*}
    \centering
    \includegraphics[width=0.99\textwidth]{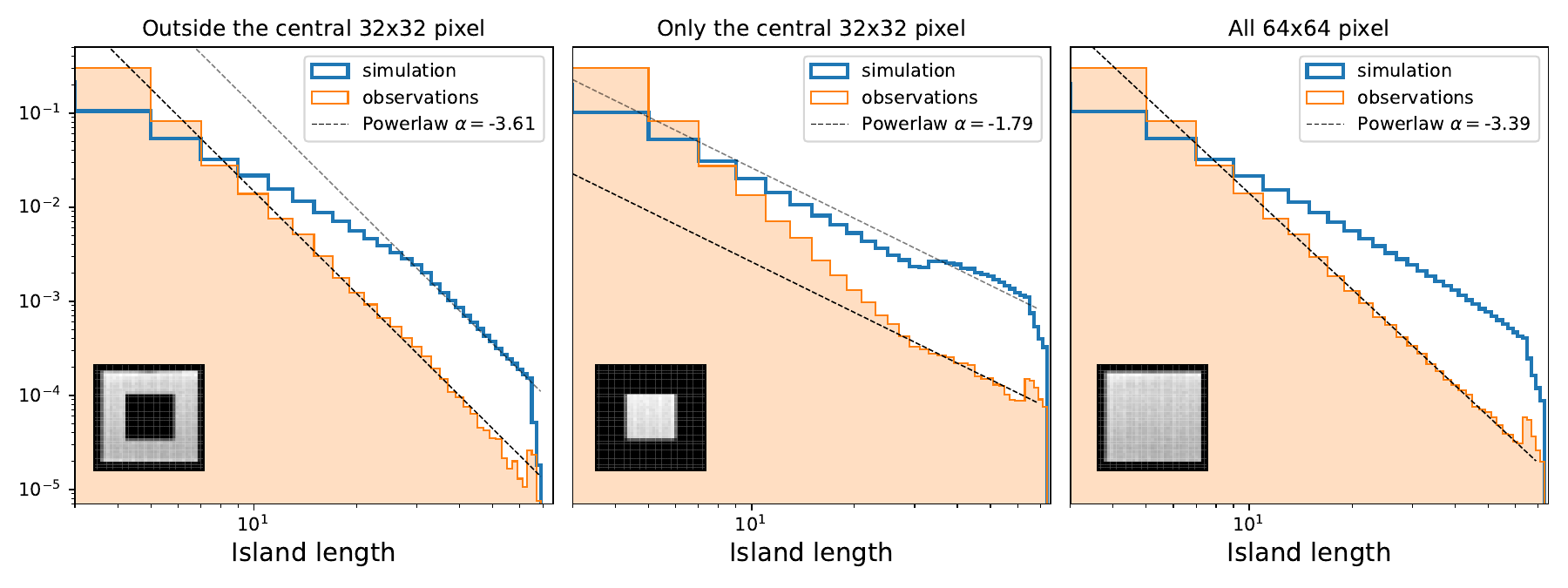}
    \caption{Island length distribution (in pixel) for three apertures: 
    Tracks with the centroid located in the central 32x32 pixels (middle panel), outside the 32x32 pixels (left panel) or anywhere on the detector (right panel). Note that these apertures only refer to the location of the island centroid, and do not necessarily exclude invalid events in the outer or inner region. 
    The apertures are shown in small insets at the bottom left of each panel. The measured distributions are shown by filled orange bars, the distributions from simple simulations in blue, and the a powerlaw fit to the longer track distributions as black dashed lines. }
    \label{fig:island_lengths}
\end{figure*}
The overall distribution of islands across the detector is best parametrized by the 2D distribution of island centroids (Fig. {\ref{fig:island_dist}} left). 
The number of islands increases with increasing RAWY, which is the detector readout direction. 
The readout of PN in SWM can be separated into 4 steps: An initial fast shift of charges before the actual integration (``clear the window''), the integration, a second fast shift to move the charges to the readout area, and the ``slow'' readout. 
Each of these steps adds a row-dependent exposure time until the charge is read out. 
The first step adds $\SI{720}{ns}$ to the row with the highest RAWY, and $\SI{46}{\mu s}$ for the lowest RAWY row. 
The second and thirds step expose all pixels uniformly, $\SI{4}{ms}$ in step 2 and $\SI{0.1}{ms}$ in step 3. Step 4 shifts charges with 1 row per $\SI{23}{\mu s}$, so the highest RAWY row has an additional $\SI{1.5}{ms}$ exposure. 
Adding up these 4 steps, we expect 34.6\% more events in the highest RAWY row compared to the lowest RAWY row. 
For more details see \citenum{Kuster1999-hr} and the \href{https://xmm-tools.cosmos.esa.int/external/xmm_user_support/documentation/uhb/}{XMM User Handbook}, Issue 2.21, Section 3.3.10. 
This is consistent with the increasing trend of islands in RAWY direction (Fig. \ref{fig:island_dist} left), where we find 33\% if we exclude the edge pixels. For the columns (RAWX direction) we do not find a linear trend across the detector, as expected. However, we see a slight edge or increase in counts in regular intervals due to the different CAMEX readout electronics, and the associated variations in amplification. 
In Case C frames we find a very similar pattern (with less statistical significance). However, a much larger fraction of the Case C frames are caused by fluorescent X-rays, when a GCR hits the electronics board and triggers a fluorescent Cu/Zn photon. This mainly occurs during the readout and not during the exposure/integration phase, since most of the Small Window is located at the Cu hole in the board. 
Therefore, the trend is expected to be even steeper for Case C frames, which is indeed what we find (156\% increase). 

The island orientation angle features no spatial dependence, and the overall distribution of angles is fairly uniform, with several peaks at values corresponding to 90 or {$\SI{45}{\deg}$}. Similarly, the spatial distribution of island lengths is also fairly uniform across the detector. However, we notice a cross-like features in the middle of the detector. This can be explained with the centroids of the longest tracks having to be located in the center of the detector. 
The distribution of island length and orientation angle is shown in Fig. {\ref{fig:island_dist}} (right). While it is clear that longer tracks are less common independently of the angle, we  notice two peaks at $-45$ and $+45$ degree and length longer than 64 pixels. The detector geometry (64x63 pixels) requires that any longer tracks must have angles other close to $\pm 45$ degree.  

When we look at the overall distribution of track lengths we find that it can be approximated by a powerlaw distribution with index $-3.4$ (see orange bars Fig. \ref{fig:island_lengths}, right panel). 
In order to isolate any edge effects we also look at two other apertures, the central 32x32 pixel, and the region outside the central 32x32 pixel. Note that these apertures only refer to the location of the island centroid, and do not necessarily exclude invalid events in the outer or inner region. 

Orange bars show the length distribution of the three apertures in Fig. \ref{fig:island_lengths}. We compare these distributions with a simple three dimensional box model of the detector, where we randomly assign particle tracks. We ensure that the simulated particle flux through all the sides of the box model is conserved, and the angular distribution of  the particle vector is uniform, i.e., an isotropic particle field. We then measure the projected length distribution, which is shown in Fig. \ref{fig:island_lengths} as blue lines. 
We find in the central 32x32 pixel area (Fig. \ref{fig:island_lengths} middle panel) the slope of the longer tracks ($\geq \SI{30}{pix}$) matches the simulated distribution (slope of about $-1.8$). However, there is steeper trend for shorter tracks, meaning we observe more short tracks than expected from a simple model.

For short-length islands none of the simulated cases (full detector, central part of detector, without central part of detector) can reproduce our observations. For the detector events excluding the central 32x32 pixels (Fig. {\ref{fig:island_lengths}} left) we find a much steeper slope of $-3.6$, while the simulations predict a flattening at the shorter track lengths. For longer island ($>25$ pixels) we find comparable slopes in Fig. {\ref{fig:island_lengths}} left and middle. We can only speculate that the excess of shorter islands is caused by a different population of particles, or secondaries in the same frame, or our detector model and isotropic particle flux assumption is not applicable for these types of particle interactions. In the future we plan to perform detailed Geant4 simulations with an accurate PN mass model to understand the length distribution.

\subsection{Spectral properties of particle background components}\label{ch:specvar}

\begin{figure*}
    \centering
    \includegraphics[width=0.99\textwidth]{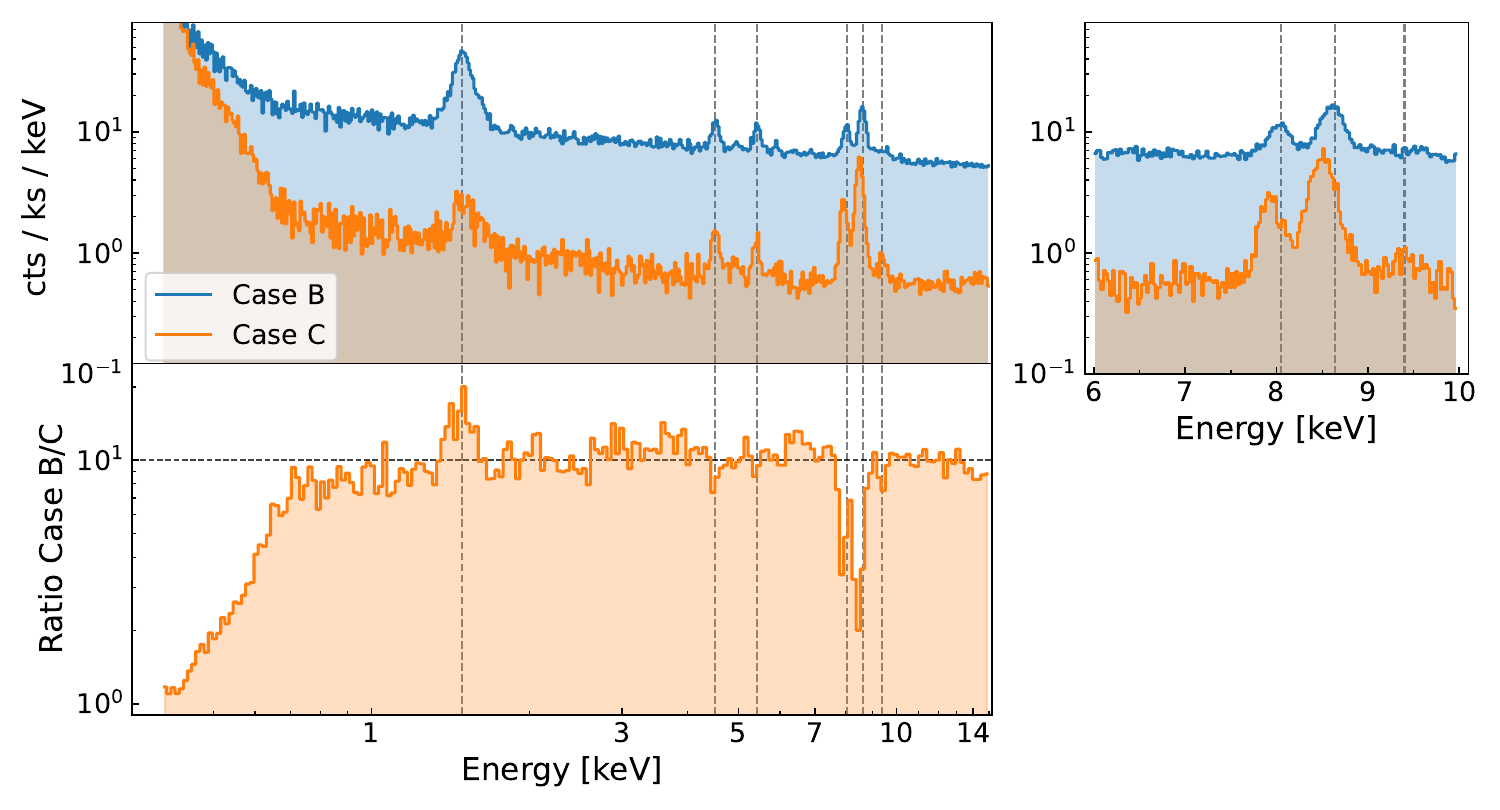}
    \caption{Top: Spectrum of valid events in Case B (blue) and Case C (orange) frames. The vertical dashed lines correspond to the spectral lines Al K$\alpha_1$ ($\SI{1.487}{keV}$), Ti K$\alpha_1$ ($\SI{4.511}{keV}$), Cr K$\alpha_1$ ($\SI{5.415}{keV}$), Cu K$\alpha_1$ ($\SI{8.048}{keV}$), Zn K$\alpha_1$ ($\SI{8.639}{keV}$), and Pt L$\alpha$ ($\SI{9.4}{keV}$). Bottom: Ratio of the Case B/C spectra to highlight differences in line strengths. }
    \label{fig:valid_spec}
\end{figure*}
\begin{figure*}
    \centering
    \includegraphics[width=0.87\textwidth]{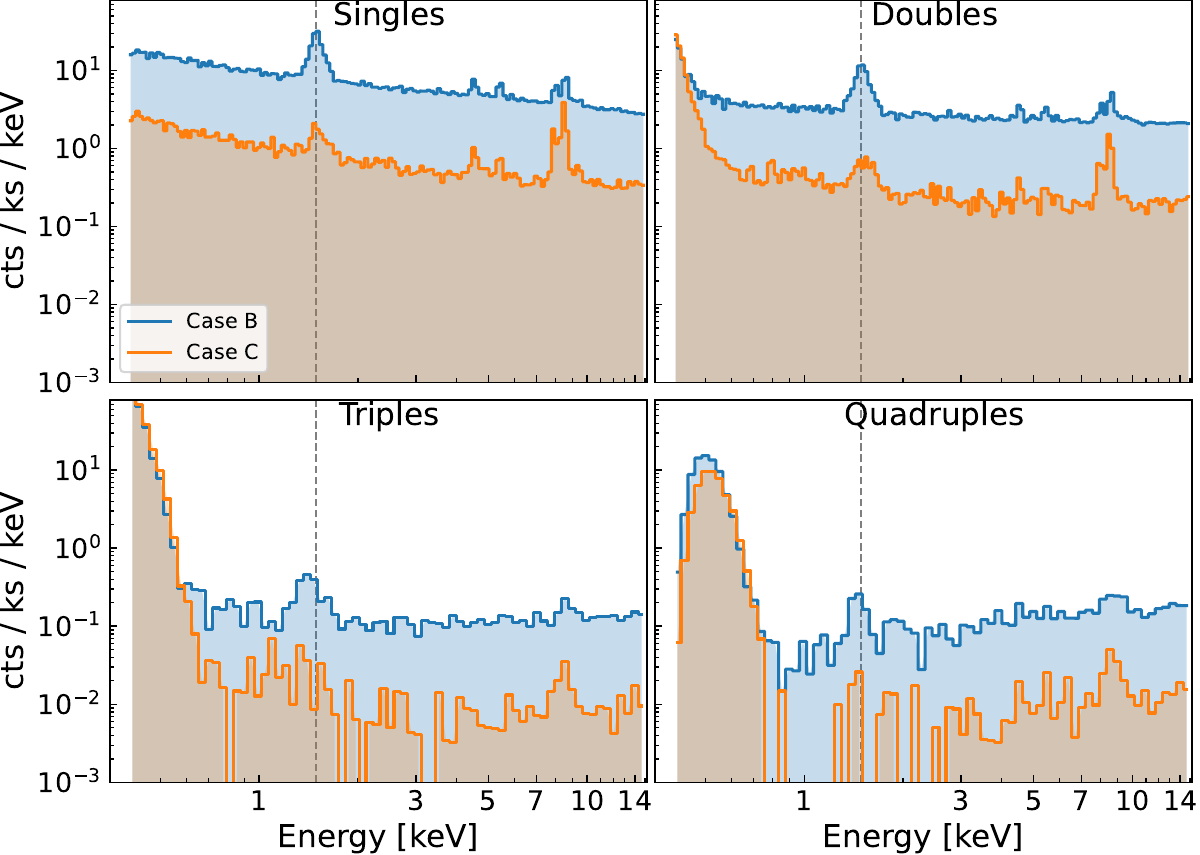}
    \caption{Spectra of valid events in Case B (blue) and Case C (orange) frames split by the event pattern distribution.}
    \label{fig:valid_spec_pattern}
\end{figure*}

We analyze the energy distribution of valid events in Case B and C frames in the PN SWM dataset with filter closed position. Figure {\ref{fig:valid_spec}} shows the spectrum of Case B (blue) and Case C (orange) valid events. The spectrum has a slowly declining continuum, with 5 clearly identified emission lines: Al K$\alpha_1$ ($\SI{1.487}{keV}$), Ti K$\alpha_1$ ($\SI{4.511}{keV}$), Cr K$\alpha_1$ ($\SI{5.415}{keV}$), Cu K$\alpha_1$ ($\SI{8.048}{keV}$), Zn K$\alpha_1$ ($\SI{8.639}{keV}$). We also see a hint of Pt L$\alpha$ ($\SI{9.4}{keV}$) in the Case C frames, which does not show up in the Case B frames, despite better statistics. 
Overall, the continuum part of the spectrum above $\SI{0.8}{keV}$ is very similar between Case B and C frames (apart from the obvious difference in normalization due to the count rate, see bottom panel in Fig. {\ref{fig:valid_spec}}). 

At the soft end below $\SI{0.6}{keV}$ we see a steep increase which mostly comes from electronic noise\cite{Lumb2002-ks,Read2003-hf,Bender2023-ue,Carter2007-ld}. It appears even steeper for Case C valid events. 
In Fig. \ref{fig:valid_spec_pattern} we find that the low energy noise becomes more and more dominant (and shifted toward higher energies) for event patterns showing more pixels above threshold, and single events are largely unaffected. 

Another striking difference between the Case B and C valid events is the shift of the Cu and Zn fluorescent lines toward slightly lower energies in Case C frames (see Fig. \ref{fig:valid_spec} right panel). This is independent of the pattern, however triples and quadruples have poor statistics to resolve the line energies. 
%We note that in filter closed spectra with the calibration source (``CalClosed'') the energies of the Fe lines agree with expectations in both Case B and C frames. 
If we look at the difference of the Case C Cu and Zn line energies from their nominal values over the lifetime of the mission, we find the the difference was largest early in the mission (up to 150\,eV in 2007) and continuously decreased to about 70\,eV in 2020. 
The PN circuit board underneath the detector has no Cu or Zn in the central region (``copper hole'', see \citenum{Freyberg2004-zn,Carter2007-ld}), which is the location of the SWM aperture. Therefore, Cu fluorescent X-ray events are very unlikely to deposit charge in the active detector region during integration. During the readout phase a fluorescent Cu or Zn photon produced by a GCR adjacent to the shifted charge region is much more likely to be detected. This also explains why these two fluorescent lines are so much brighter in Case C than in Case B frames (relative to the continuum). The few fluorescent Cu and Zn photons in Case B frames are likely produced during the integration phase from a GCR that misses the aperture, while the fluorescent photon scatters into it. During readout however, the GCRs can easily produce fluorescent Cu and Zn photons that are detected, but end up in a frame with a particle track, as Case C frames. The readout is likely going to affect the detected photon energy, which systematically shifts these lines in Case C frames. 
For Case C frames with a Cu/Zn X-ray the single pattern fraction is much higher than for Case B Cu/Zn photons (73.9\% vs. 60.5\%, with 1.6\% and 1.0\% statistical uncertainty, respectively). This might explain the shift toward lower energies of the Cu/Zn lines in Case Cs: The Cu layer is physically very close to the detector, and fluorescent photons will therefore be also close to any particle track. If the charge in the detector from the Cu/Zn photons is split over multiple pixels (pattern), some of it might be confused with the nearby particle track and is not correctly added to the X-ray-like event. This will reduce the charge in Case Cs, but not in Case Bs, which do not contain a particle track. 

\begin{figure*}
    \centering
    \includegraphics[width=0.99\textwidth]{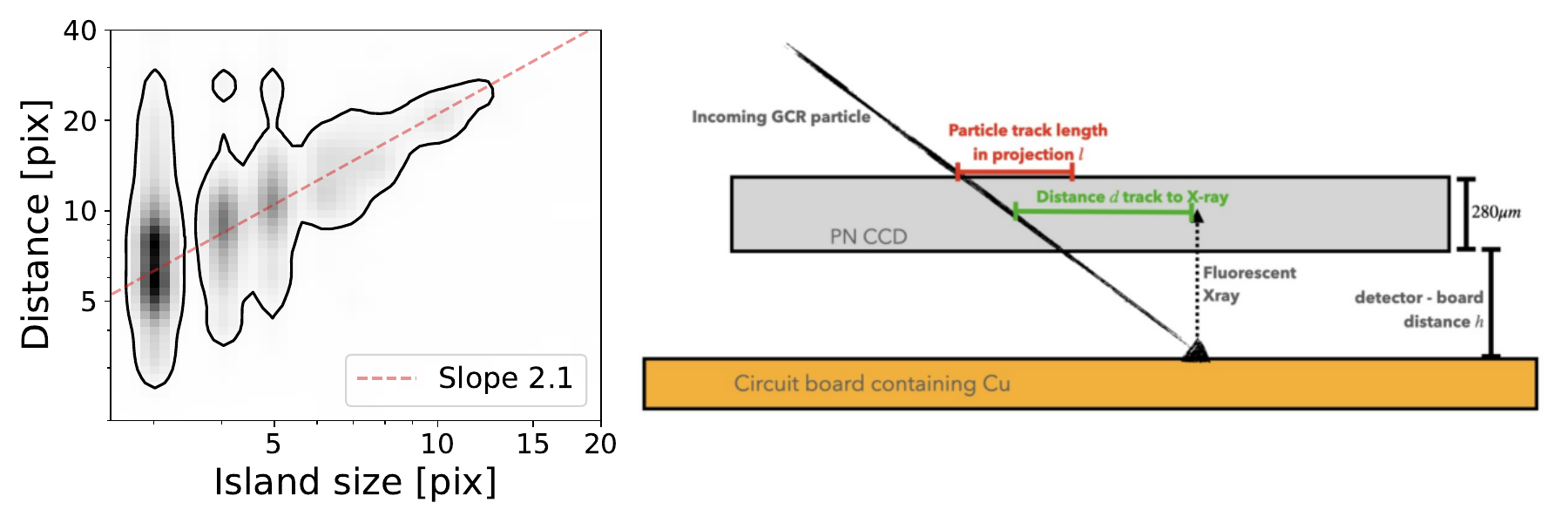}
    \caption{Left: Fluorescent Cu X-rays in Case C frames show a correlation between the length of the associated particle track in the frame $l$, and the distance between X-rays and track. Right: Illustration showing the path of a GCR through the detector, leaving a track (red), and triggering the emission of a Cu X-ray from the circuit board underneath. The distance between the board and the CCD $h$ can be derived with simple geometric assumption. }
    \label{fig:size_dist}
\end{figure*}
In order to verify our picture of the Case C fluorescent Cu events, we look at these events in a bit more detail: From the 775 Case C frames with a valid event energy consistent with Cu, we correlate the particle track length with the distance between the track and the X-ray event in each frame. 
We compare the smoothed distribution with a slope 2.1 in Fig. {\ref{fig:size_dist}} (left).  
The larger the angle of the incoming GCR, the longer the track and the further we expect the fluorescent X-ray to be detected from the GCR detection. 
We illustrate the geometry in Fig. {\ref{fig:size_dist}} (right), where one can easily conclude the following identity:
\begin{equation}
    h = \SI{280}{\mu m}~ \frac{d}{l},
\end{equation}
where $\frac{d}{l}$ is the slope in Fig. {\ref{fig:size_dist}} (left). We can therefore derive the separation between the detector and the board. We find a separation of about $h \approx 2.1 \times \SI{280}{\mu m} -  \SI{140}{\mu m} \approx  \SI{450}{\mu m}$, which is a reasonable result considering that the actual distance is $\SI{250}{\mu m}$ ({\citenum{Struder2003-zm}}), and the pixel thickness (depletion depth) is $\SI{280}{\mu m}$.

\subsection{Temporal variability of the particle background}\label{ch:timevar}
The time variability on various scales remains a significant source of systematic uncertainty in the NXB (e.g., \citenum{Bender2023-ue}). 
A rapidly changing particle rate and/or spectrum will render any blank sky background subtraction very difficult, and likely insufficient for future missions such as Athena WFI, which require a precise knowledge of the background\cite{Rau2013-qp}. Instead a deeper understanding of the particle background variability and its dependencies is required to develop models and mitigation methods. In the following we analyze the lightcurves of the particle interactions with the PN and the Chandra HRC anti-coincidence shield. The latter has negligible statistical uncertainty and (almost) continuous sampling with extremely high time resolution for 20 years. 

\subsubsection{PN Small Window Mode}\label{ch:PNSWM}
\begin{figure*}
    \centering
    \includegraphics[width=0.99\textwidth]{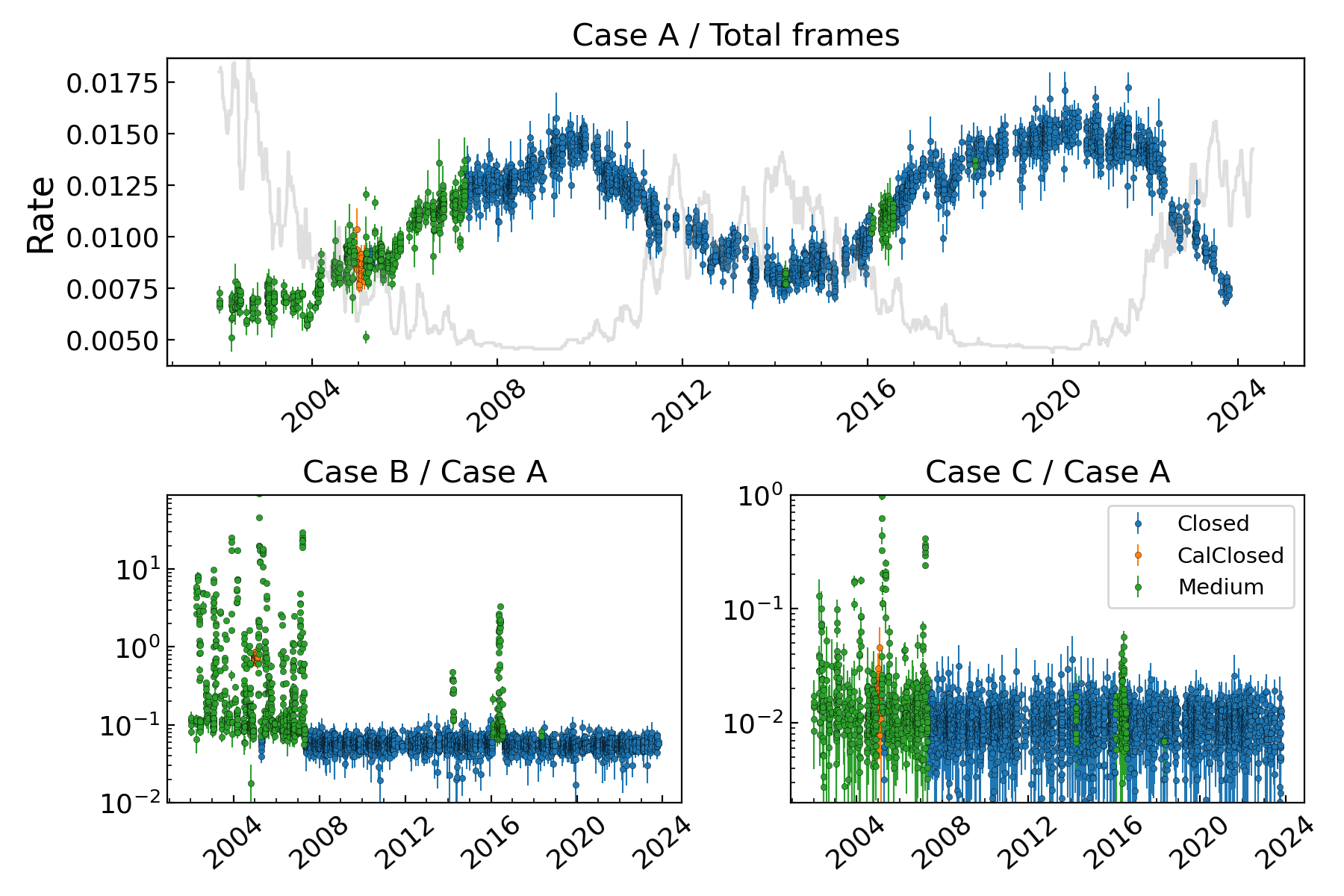}
    \caption{Lightcurves of the PN SWM slew observations with 10 minute time resolution. Top: Case A (only invalid events in frame) frame rate. We also show the average number of sunspots in grey (arbitrary y scale) as provided by WDC-SILSO, Royal Observatory of Belgium, Brussels. Bottom left: Ratio of Case B to Case A frame rate. Bottom right: Case C frames per Case A frame. The color in all panels refers to the filter position (closed: blue, open/medium: green, closed with calibration source: orange).}
    \label{fig:lightcurves}
\end{figure*}

We select our previously analyzed slew observations in SWM, which have been taken either in FWC mode, medium filter, or CalClosed mode. 
On average we find that about 1 to 2 percent of all frames contain invalid events. Therefore, the chance of randomly detecting two independent GCRs in a single frame is $<\SI{0.04}{\%}$, and therefore negligible. 
The number of invalid events (i.e., the total number of pixels in particle tracks) is not a good measure for the particle background rate, as, depending on geometry, tracks appear longer and sometimes events from secondary particles are also detected. We can therefore assume that invalid events in the same frame are not independent. 
Also the number of islands/tracks is not a good measure of the primary particle rate, as a single primary can trigger a ``shower'' of secondary events which might be detected as invalid events in separate islands within the same frame (this happens in $\sim 5\%$ of the Case A frames). Therefore, the best quantity to use is the number of frames with invalid events, as this is the least noisy tracer of the particle rate.

We calculate the rate of invalid events in 10\,min time bins, with each bin having typically 1000 invalid frames (about $\SI{3.2}{\%}$ expected statistical uncertainty). 
We actually measure an RMS scatter from bin to bin of $\SI{3.4}{\%}$, very close to our expectations, therefore, showing that on 10\,min intervals the statistical uncertainty dominates over systematic changes within the same period.
Figure \ref{fig:lightcurves} (top panel) shows the Case A frame rate over more than 20 years (almost two full solar cycles). On timescales of years the lightcurve clearly follows the solar cycle, as initially suggested by \citenum{Gleeson1968-ik}. For comparison we overplot the number of sunspots in Fig. \ref{fig:lightcurves} (grey line in top panel), which anti-correlates with the particle background rate. The scatter in the Case A rates from closed and medium filter observations (blue and green points, respectively) is determined with respect to a smooth spline that we fit to the lightcurve. We find that FWC observations have a scatter of 4.8\%, while medium filter observations have 5.3\% scatter. However, the two sided KS-test shows that the two distributions, residuals from closed and medium filter observations, are the same (p-value 0.89). We can therefore conclude that we see no indications that the Case A frames, even for medium filter observations, are affected by sky X-rays. Case A frames trace the particle background level and any method that utilizes the particle tracks in Case A or C frames to estimate or mitigate the X-ray particle background can be applied to sky observations as well.

The Case B frames behave very differently. Figure \ref{fig:lightcurves} (bottom left panel) shows the Case B frame rate normalized by the Case A rate. For FWC observations (blue) the ratio is almost constant, showing that there is no other source of Case B events. 
\begin{figure*}
    \centering
    \includegraphics[width=0.99\textwidth]{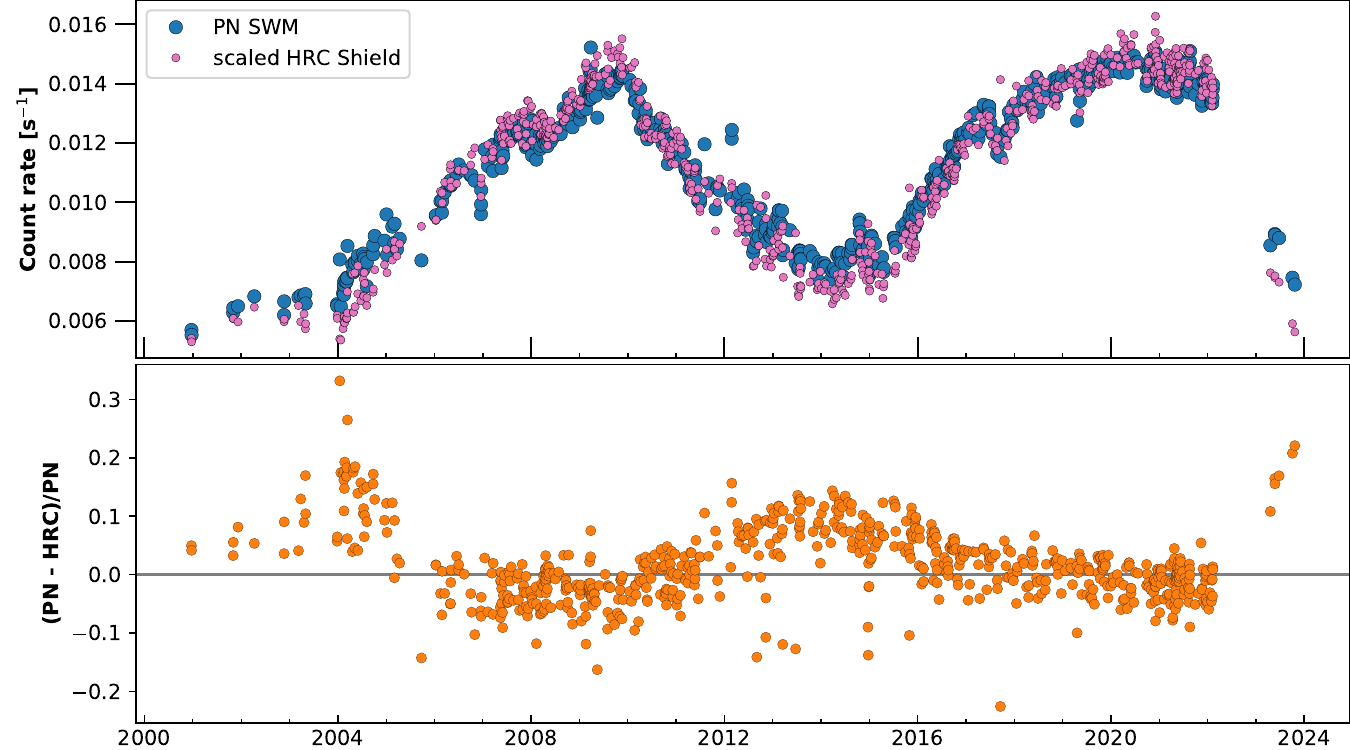}
    \caption{Comparison of the Chandra HRC shield lightcurve and the XMM PN SWM Case A frame rates. Top panel: Direct comparison of the rates, where the HRC shield data (purple, typically several thousand counts per second) are rescaled by a constant factor to overplot with the much smaller, and also rescaled with a time dependent function to correct for the loss of PMT sensitivity. Bottom panel: Relative residuals between PN SWM Case A frames and HRC shield count rates.}
    \label{fig:hrc_scaling}
\end{figure*}
For Medium filter observations (green) we find a higher baseline level due to celestial X-rays, and also over two orders of magnitude of scatter. For these variations to originate from bright X-ray sources that happen to lie on the slew path, they would need to be extremely bright: The XMM-Newton slew rate is about $\SI{90}{\deg}$ per minute, which means that a source is only within the SWM aperture for about 4 seconds. The typical quiescent count in Case B frames is $\num{e-3}$, which means about $\SI{0.18}{ct\,s^{-1}}$, or 650 counts in a typical 1\,h observation. Therefore, a bright source, which is visible for about 4\,s, will have to contribute about $\SI{64000}{cts}$ in order to increase the overall count rate by two orders of magnitude. This means the source needs to have a flux of at least $\SI{3e-8}{erg\,s^{-1}\,cm^2}$, which is more than the Crab SNR. We can safely assume that XMM-Newton does not regularly slew across Crab-like sources, and exclude the possibility that celestial X-rays cause the spikes seen in medium filter slew observations. The only other explanation is soft protons, that get partly focused by the XMM optics and cause X-ray like patterns and energy signatures\cite{Aschenbach2007-fk,Fioretti2017-ac}. 
We verified this by analyzing over 2700 slew observations in PN Full Frame mode. The Case B frames can be easily identified in these observations, but unlike the SWM we are unable to measure the Case A frames. 
Since we know the trend with time from the SWM observations (Fig. \ref{fig:lightcurves} top panel), we can interpolate to obtain an approximate Case B/A ratio for FF observations. When comparing these rates with the values from the Fin/Fout test\cite{De_Luca2004-le} we find consistency, meaning the same observation that are flagged as soft proton contaminated, have a high Case B/A rate. However, since there is some proton contamination in the out-FoV region of the PN\cite{Marelli2021-jf} the Fin/Fout ratio saturates at a certain soft proton flux, while the Case B/A ratio keeps rising and appears to provide a more stable measurement of soft proton contamination in the PN, at least for our slew observations without bright targets. 
We investigate the properties of soft protons in section \ref{ch:softprotons} in more detail. 

Lastly, the Case C frames show a very similar trend to the Case A frames (Fig. \ref{fig:lightcurves} bottom right). The closed observations follow the Case A frames (Case C scatter 20\%). Observations with the medium filter have Case C/A ratios that are also mostly constant, with some outliers, and their scatter is significantly larger (37\%). However, apart from a few outliers, the Case C frames are mostly related to the Case A, and mainly particle induced. 

\subsubsection{Chandra HRC shield}
The HRC shield count rates shows a very similar trend to the PN SWM Case A rates, where we clearly see the solar maxima with lower count rates around 2003 and 2014, and increased particle rates peaking in 2010 and 2020 (Fig. \ref{fig:hrc_scaling} purple data points in top panel). However, while the two datasets are highly consistent early on, the HRC shield rates appear to decline with time relative to the PN SWM (this effect is already corrected for in Fig. \ref{fig:hrc_scaling}).  This is expected due to the known increase in opacity in the scintillator and the loss of sensitivity over time of photomultiplier tubes (PMTs, e.g., \citenum{Hum2009-dz}). The PN SWM Case A frame rates can be used as a reference to correct the HRC shield rates. We implement a linear, time dependent function to match the HRC shield and PN SWM rates, while making sure to only include rates where both HRC shield and PN SWM data are available. For this scaling we use 24h time bins. Figure \ref{fig:hrc_scaling} (top panel) shows the PN SWM Case A measurements (blue) together with the scaled HRC shield count rates (purple). 
The correction terms account for a linear decrease in HRC shield sensitivity of 2.9\% per year, and residuals are typically below 10\% (see bottom panel in Fig. \ref{fig:hrc_scaling}). 
The verification of the HRC shield data's consistency with PN SWM Case A rates highlights that two instruments on different satellites in slightly different orbits trace the same particle rate, which allows the application of conclusions from the high statistics HRC shield data to PN SWM data as well.

\begin{figure*}
    \centering
    \includegraphics[trim=25px 0px 10px 0px,clip,width=0.99\textwidth]{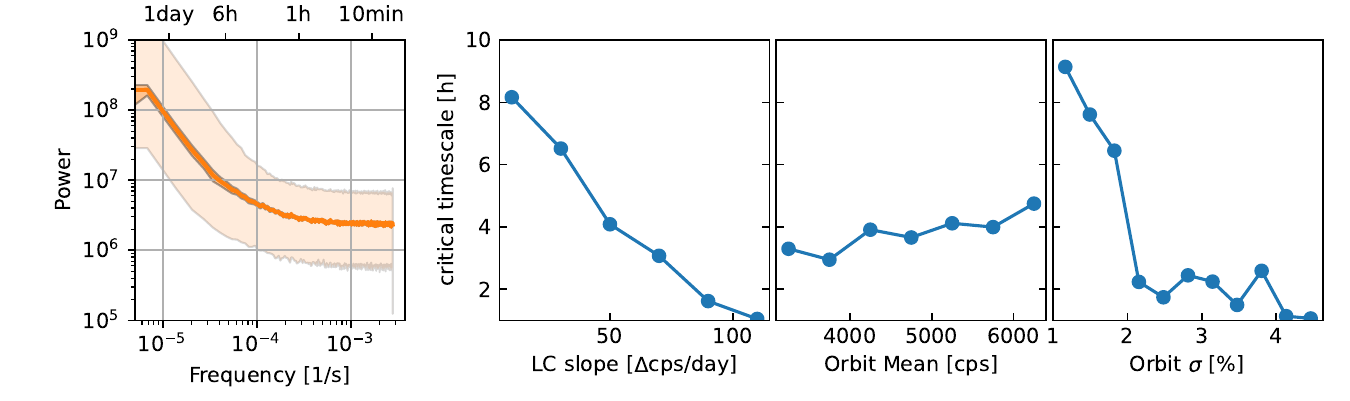}
    \caption{Left: Power spectrum of the median HRC shield lightcurve, showing significant power in the particle background variability above 1\,hour. The width of the dark orange bar represents the standard error, while the light orange shaded area shows the scatter of all power spectra.  
    Three right panels: Dependence of the critical timescale (power increase by factor of 2) on orbital parameters (lightcurve slope, mean rate in each orbit, scatter in each orbit). A higher critical timescale represents a more quiescent particle background. }
    \label{fig:power}
\end{figure*}

However, the residuals on long timescales (years) appear to follow the solar cycle (PN higher at solar maximum, lower panel of Fig. \ref{fig:hrc_scaling}). 
The reason is likely a time variable spectrum of the GCRs (e.g., \citenum{Yamashita2008-sc}) that causes the distinct measurements in various instruments depending on the instruments energy sensitivity. 
Hardness ratios from Chandra ACIS in stowed position of the back-illuminated chips showed a similar variability{\cite{Suzuki2021-ig}}, which also leads to the conclusion of changes in the CR properties. 
The HRC shield is most sensitive to particles with energies of tens of MeV, while the PN SWM Case A events are mostly induced by GeV protons. 
However, the scatter of these residuals also varies: From 2002 to 2005 the scatter was $5.7\%$, and during the next solar maximum, 2012 to 2016, it was $5.5\%$. In between the solar maxima, from 2006 to 2012, and from 2016 to 2022, the scatter was significantly lower (3.2\% and 3.0\%, respectively). We also reanalyzed these data with shorter time binning, instead of our default 24h, which results in higher values for the scatter. For example, for the 2006 to 2012 period, we get 3.6 (3.4, 3.2, 3.1, 2.9)\% for 1, (6, 24, 48, 96) hour time bins, respectively. The trends are similar for the other intervals. 
We note that if the statistical uncertainty was the dominant source of scatter, we should expect much smaller values even for the 1h time binning, since the statistical uncertainty is typically 0.02\% for HRC shield, and 1.2\% for PN SWM. 
On the one hand, this clearly indicates that while the differences between HRC shield and PN SWM rates follow a normal distribution, they are systematic and time dependent. On the other hand, these residuals are more pronounced when comparing shorter time intervals, which emphasizes the importance to have a reliable background measurement close in time. 
It is possible that variations due to location (local magnetic field) can explain part of the residuals. A collection of simultaneous particle rate measurements in different orbits (other than XMM-Newton and Chandra) will help to quantify the orbital contribution to the residuals.

\citenum{Sarkar2024-rx} found a 6-day lag between the lightcurves of AMS and Chandra/XMM during the time period of  2016 and 2017. We searched for a similar time delay between lightcurves, making use of our superior time resolution. 
We re-extracted lightcurves on 30\,min intervals, and after applying the previously calculated correction factor for the loss of PMT sensitivity, we utilize a timing analysis with a discrete correlation function. We find no time delays on any timescales of a less than 10 days.
However, we find the peaks in the correlation function at 27 days time difference, and multiples 27 days. These are related to the (average) solar rotation timescale of sunspots\cite{Willson1988-op,Beck2000-oy,Reuveni2009-zk}.

The wealth of information contained in the Chandra HRC shield dataset allows us to analyze a power spectrum of particle background variability. This utilizes the exceptional statistical power and extremely high time resolution with (almost) continuous observations for over 20 years. 
However, the elliptical orbit of Chandra passes through the Van Allen belts, which temporarily interrupts all science observations, including the operation of the HRC shield\cite{Virani2000-ze}. Therefore, the total orbital period of about 65\,h permits continuous observations for only about 55\,h. We derive a lightcurve of each Chandra orbit from the HRC shield rates, and quantify the orbital mean rate, its standard deviation, and the linear slope of the time dependence. 
We exclude the 3\% of the most active orbits, where the scatter of the rates is at least 5\%. For each orbital lightcurve we derive the power spectrum (PS) from the Fast Fourier Transformation (FFT), and stack all PS to derive the median PS. Figure \ref{fig:power} (left) shows the median power spectrum and its $1 \sigma$ scatter as the lightly shaded orange region, while the darker orange shade represents the standard error, which is relatively small due to the large number of orbits. 
The offset at the highest frequencies is due to the normalization of the FFT. 
The orbit-to-orbit scatter does not change with orbital properties, such as mean rate, slope or rate standard deviation. 
We can clearly see that there is more power on longer timescales, while on times below 1\,h there is no significant power (variation consistent with the range of standard error). The longest timescale we can probe with our analysis is 2.3\,days, the duration of the Chandra orbit outside the radiation belts. 
The shape of the power spectrum is well fit by a powerlaw plus constant, which allows us to define a critical timescale, at which the power reaches twice the constant baseline at high frequencies. The longer this critical timescale is, the less time variable is the particle background, and it can be assumed to be almost constant on timescales shorter than the critical timescale. For our default power spectrum, we derive a critical timescale of 3.9\,h. 
In the three right panels of Figure \ref{fig:power} we test the dependence of this critical timescale on the orbital properties: For a strong linear trend meaning either a decreasing or increasing lightcurve, we find significantly shorter critical timescales (``slope'', which is measured as the absolute value), while, for the flattest lightcurves, critical timescales of up to 8\,h are found. The mean rate in each orbit, mostly dependent on the solar cycle, is weakly correlated with the critical timescale. During solar minimum, fewer sunspots are present, but the particle background level is higher. However, fewer sunspots also imply a more stable background, and therefore longer critical timescales. Lastly, the scatter of the particle rate during an orbit is correlated very clearly with the critical timescale. Less than 2\% scatter in the rate translates into critical timescales longer than 6\,h, while the critical timescale is between 1 and 3 hours for orbits with greater scatter. 

\subsection{Soft proton contamination}\label{ch:softprotons}
\begin{figure}[tbp]
    \centering
    \includegraphics[width=0.6\textwidth]{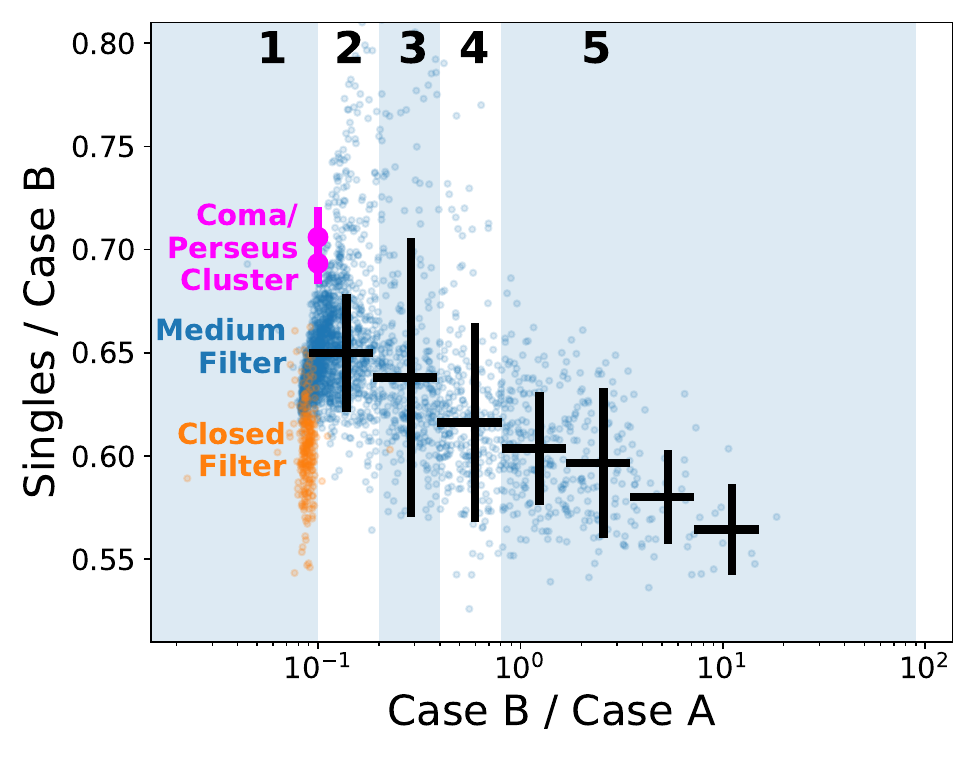}
    \caption{Dependence of the single pattern fraction on the soft proton indicator, the Case B/A frame ratio. Blue points mark medium filter, full frame observations in slew mode (no science target), orange point show the distribution of closed filter observations, the two magenta point show the location of two very bright, extended X-ray sources. Black crosses are averages of the blue points. We define 5 levels of SP proton contamination, indicated by the labels 1 to 5.}
    \label{fig:casab}
\end{figure}
We have analyzed the PN SWM lightcurves in section \ref{ch:PNSWM}, where we concluded that the Case B/A frame ratio (frames with only valid events divided by only invalid event frames) can be used as a reliable soft proton contamination measure. 
As laid out there, this indicator can also be applied to all full frame (FF) observations in slew mode, since the Case A rate can be interpolated from Figure \ref{fig:lightcurves}. 
In order to characterize the soft proton events further, we look at the fraction of single pattern events in these FF observations (with medium filter) and compare it with other datasets. Figure \ref{fig:casab} shows the singles fractions versus the Case B/A estimator for the medium filter FF slew observation (blue). We categorize the soft proton contamination into 5 levels (numbers 1 through 5 in Fig. \ref{fig:casab}), and also show binned averages of the blue points in black. Observations free of soft protons flares have singles fractions around 65\%, while the most extremely contaminated observations reach 55\%. These averages clearly show that soft protons typically create fewer singles, meaning the  charge is distributed over a larger number of pixels. 
The GCR particle background events (shown in orange) have a similar singles fraction to observations with strong soft proton contamination. A typical bright, diffuse or extended celestial X-ray source (in this case the bright clusters of galaxies Coma and Perseus) have higher singles fractions, around 70\%. 
Since all of the slew observations are exposed to the diffuse X-ray foreground emission from the Galaxy, the Local Hot Bubble, and the Cosmic X-ray Background, but also to the particle background, it seems plausible that observations with few soft protons have singles fractions between the clusters and the closed filter observations. 

\begin{figure}[tbp]
    \centering
    \includegraphics[width=0.49\textwidth]{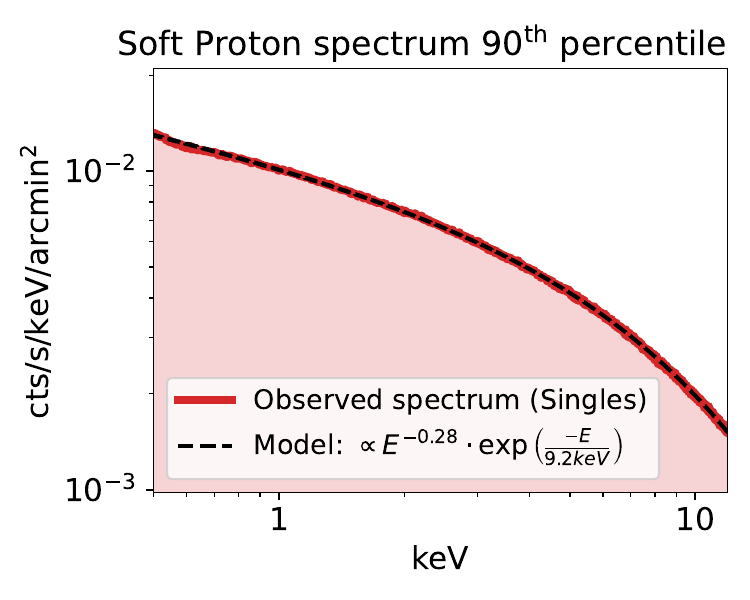}
    \caption{Soft proton spectrum derived from observations within the 85th and 95th percentile of Case B/A ratios. The spectrum is ``background subtracted'', meaning that an average spectrum with low Case B/A spectrum has been subtracted. The black dashed line shows a fit of a powerlaw function with exponential cutoff.}
    \label{fig:sp_spectrum}
\end{figure}
After we have classified all the FF slew observations in the medium filter we can select a subset with high soft proton contamination and analyze the spectrum. 
We note that previous results{\cite{Kuntz2008-mn}} for the MOS cameras on XMM-Newton indicate a time variability of the spectrum of the soft proton induced X-ray background. While we see variations in the PN soft proton spectrum, we do not quantify this here. Instead, we concentrate on the average spectrum to be compared with simulations at a later stage. We focus exclusively on single events in the following, since they have the best statistics, and other patterns might have different spectra. 
In Fig. \ref{fig:sp_spectrum} we show the average spectrum of exposures around the \nth{90} percentile of the Case B/A ratios after subtracting an average quiescent spectrum from it.
The subtraction appears to be satisfactory since no fluorescent lines are present in the data (or absorption features from over-subtraction). The spectrum is well described by a powerlaw (index $\alpha  = 0.28$) with an exponential cutoff at around 9\,keV. 
We note that this observed spectrum can motivate and verify detailed Geant4 simlations\cite{Fioretti2021-tm}, which will in turn allow a better modeling of the soft proton component in observations through tailored response matrices.

\section{Utilizing self anti-coincidence to reduce the particle background}\label{ch:sac}
Case C frames contain both, a particle track and an X-ray-like event. This provides additional information to reduce the particle background. Foremost, we can analyze the distance between a track and an X-ray event that occur in the same frame. Due to the short frame time of the SWM ($\SI{5.6718}{ms}$) there is a high chance that these events (valid and invalid in Case C) are correlated, meaning the particle also created the X-ray event (see also \citenum{Miller2022-yy}).

We know that all frames with invalid events, $I$, are Case A and C frames, $I = A + C$, while the valid event  frames are Case B and C frames, $X = B + C$. 
If we assume for now, that the particle and X-ray events in Case C frames are random and not correlated, we can derive the expected number of C frames,
\begin{equation}\label{eq:sac_scaling}
    C = \frac{I \cdot X}{T} = \frac{(A+C)\cdot (B+C)}{T}~,
\end{equation}
where $T$ is the total number of frames. 
The result can be compared to the actual number of Case C frames to provide insight into the absolute/integrated probability of correlated events in Case C frames. We note that a small number of frames can be misidentified, when a valid event is within 1 or 2 pixels of an invalid event in the same frame. We estimated the effect of this on the probability of correlated events in Case C frames to be less than 1 percent. 

We can thus utilize the distance between particle tracks and X-ray events to derive a probability that the X-ray event was particle induced. For this we compare a random distribution of distances with the observed distance distribution, and normalize the result by the expected random number of Case C frames. 

\subsection{Random distribution of distances}
\begin{figure*}[tbp]
    \centering
    \includegraphics[width=0.99\textwidth]{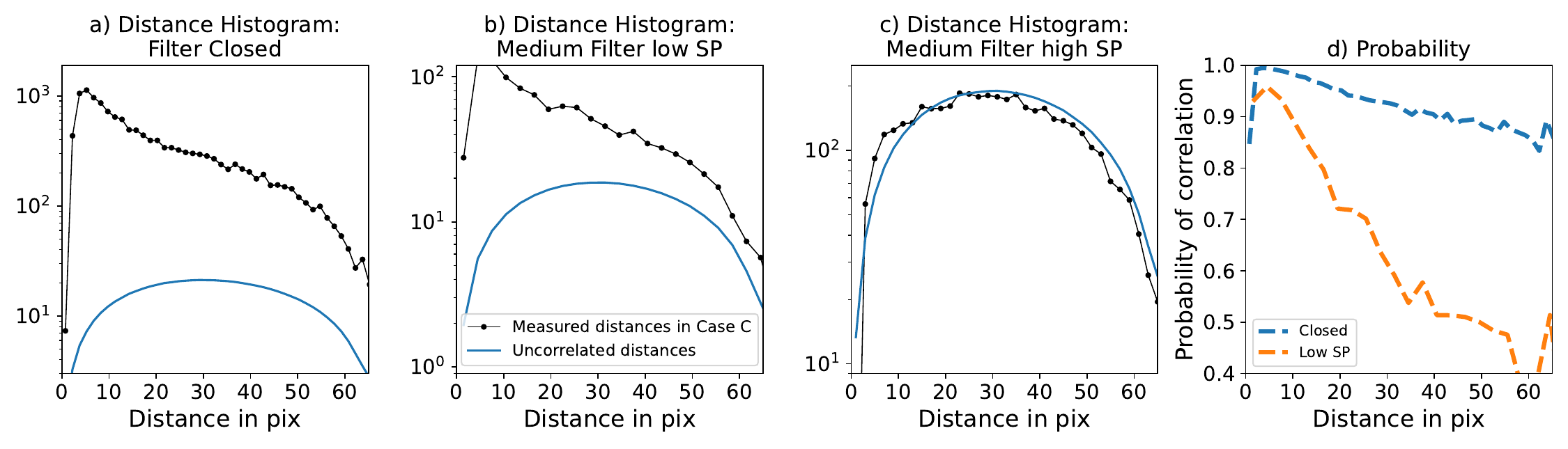}
    \caption{Assessing the capabilities of self anti-coincidence (SAC). Panels a-c show the distribution of distances (black lines) between valid and invalid events in Case C frames, with only FWC observations in a), medium filter slew observations with low SP contamination in b), and, in c), for high SP contamination. The blue lines show the expected distribution for uncorrelated events.Panel d) shows the probability of an X-ray event at a given separation to the particle track, being induced by a particle  (blue: closed filter observations, orange: medium filter with low SP). High soft proton contamination is not shown in d) as it is consistent with 0. }
    \label{fig:sac}
\end{figure*}
\citenum{Bulbul2020-zf} used an idealized random distribution for points on a disk of radius $R$ to derive the probability function.  
However, assuming a disk instead of rectangular detector will introduce a small bias. 
In Section \ref{ch:spatialvar} we have derived the 2D distribution of events across the detector, which is nonuniform and should also be taken into account. 
We now derive empirically the distribution of random pair separations from the observed particle distribution on the detector (see Fig. \ref{fig:island_dist} top left). 
We also take into account the distribution of valid events (as observed in Case B frames), which increases from the lowest to highest RAW Y column by $\sim 30\%$. 

In order to compare the observed distribution of distances between X-ray and particle events in Case C frames, we use Eq. (\ref{eq:sac_scaling}) to normalize our empirically determined random model. By comparing these distributions, the observed one and the random expectation, we can quantify at a given distance how likely an X-ray-like event is to be associated with a GCR. 
We show the two distributions (observed in black, random in blue) in Fig. \ref{fig:sac}, for (a) slew observations with the filter in the closed position, which means no sky X-rays are detected, for (b) slew observations with the medium filter and a low Case B/A ratio, meaning little soft proton contamination, and (c) medium filter observation with a very high soft proton (SP) contamination. 
We clearly see that shorter distances are generally more likely than larger distances, and that for FWC observations, the distribution of random coincidences underestimates the observed number of Case C events by at least an order of magnitude at all distances. 
We note that for distances of 0 or very few pixels we are unable to distinguish the X-ray event from the particle track. Therefore, extremely short distances ($<3$ pixels) are not reliable in Fig. {\ref{fig:sac}}. 
For the medium filter and low SP case, the shorter distances are highly over-represented with respect to the random expectation, while larger distances are almost in agreement. 
Finally, when there is intense soft proton contamination the agreement between the two curves is good, meaning almost none of the valid events in Case C frames are due to particle background events. 
These results are also reflected in panel (d) of Fig. \ref{fig:sac}, which shows the probability that an X-ray event at a given distance from the particle track is correlated with this track. 
Note that this is not a probability density function or probability distribution, and therefore the curves in Fig. \ref{fig:sac} (d) do not add or integrated to 1. 
The FWC data are shown in blue, decreasing from about 100\% at the shortest distances to about 88\% at 64\,pixel distance. This means, 96\% of all FWC X-ray-like events in Case C frames are induced by a GCR. Uncertainties in finding the center for the particle track, and noise induced X-ray-like events cause the majority of the other 4\% of events. The medium filter events in Fig. \ref{fig:sac} (d, orange curve) reach close to 100\% at the shortest distances, but decrease much more steeply to about 47\% at 64\,pixel, making about 78\% of all medium filter Case C X-ray events GCR induced. 
Note that we do not show the curve for the high soft proton observations, since it is essentially consistent with all events being unrelated to particle events. We also note that the statistics for very short distances ($\leq \SI{3}{pixel}$) are very poor and it becomes hard to clearly distinguish X-ray events from particle tracks.

\subsection{Optimized self anti-coincidence}
\begin{figure*}[tbp]
    \centering
    \includegraphics[width=0.99\textwidth]{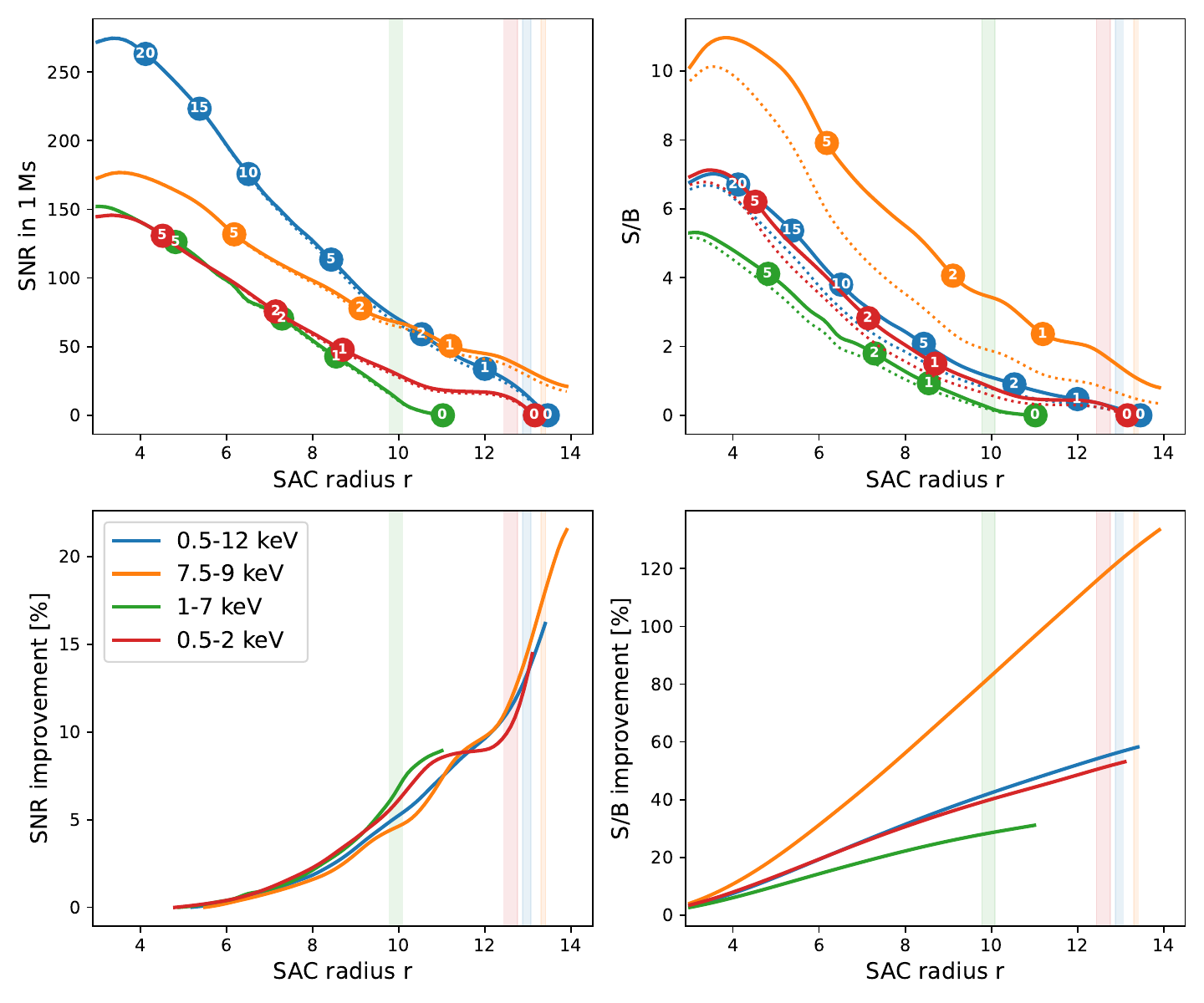}
    \caption{Demonstration of the optimized self anti-coincidence (SAC) method. In all panels the x-axis is the optimized SAC radius for a given source brightness ($r$ as a function of the source surface brightness). The top left (right) panel shows the signal-to-noise, SNR,  (source to background ratio, S/B) as a function of the SAC exclusion radius, assuming a uniform source with the optimal brightness, and a 1\,Ms exposure. The corresponding surface brightness values $\bar S$ are indicated in the circles in units of $\si{cts\,pix^{-1}\,Ms^{-1}}$. The four different colors correspond to the energy bands of valid events listed in the legend below. Dotted lines refer to the SNR (S/B ratio) without SAC. The bottom panels show the improvement in \% on the SNR and S/B. The shaded vertical bars indicate the median background level of blank sky observations (low SP case).}
    \label{fig:sac_improvement}
\end{figure*}
Here we consider how to optimize the signal-to-noise ratio for the Class C frames.
We assume the signal-to-noise ratio to be
\begin{equation}
    {\rm SNR} = \frac{S}{\sqrt{S+B}}~,
\end{equation}
where $S$ and $B$ are the source and background counts. 
For a point source, the SNR is generally optimized by considering only events that lie close to the source position.  However, the best strategy is different for extended sources.
Our goal is to maximize the SNR when a circle of radius $r$ is excluded around each particle event. 
We can determine $S$ and $B$ as functions of $r$ simply by excluding events that lie too close to an invalid event, assuming that the source emission is uniform,
\begin{equation}\label{eq:source}
S(r) = \bar S (A_F - \pi r^2)~,
\end{equation}
where $\bar S$ is the average count per unit area, and $A_F$ is the field of view area.  For $B(r)$, we determine the total number of GCR induced X-ray events lying further than $r$ from particle tracks. We empirically determine this by integrating the FWC distance distribution {\ref{fig:sac}} from $r$ to a large radius. 
The optimal $r$ can be found by maximizing the SNR, i.e. by locating the zeroes of $d\,{\rm SNR} / d\,r$, which yields 
\begin{equation}\label{eq:optimal_sac}
    \bar S = -\frac{2 {B}}{A_F - \pi r^2} - \frac{\frac{d {B}}{d{\rm r}}}{2 \pi r}  ~.
\end{equation}
This form relates surface brightness of the source to the optimal exclusion radius, in terms of the spatial distribution of the X-ray events induced by GCRs. It can be inverted to determine the optimal exclusion radius for a given source brightness.
Since pixels are discrete, a GCR induced X-ray event cannot lie less than one pixel away from a track.  Thus a meaningful $B(r)$ can only be determined for $r >1$ pixel and meaningful values for $\frac{d{B}}{d r}$ can only be determined for $r \ge 5$ pixel.  We find that $\bar S$ given by equation (\ref{eq:optimal_sac}) increases for small $r$ to a maximum of $\bar S_{\rm max}$, before decreasing and, eventually, going negative for large $r$.  Thus, this optimization is only feasible for surface brightnesses smaller than $\bar S_{\rm max}$.
We note that $\bar S_{\rm max}$ is proportional to the exposure time through the dependency on $B$. 
\begin{figure*}[tbp]
    \centering
    \includegraphics[width=0.99\textwidth]{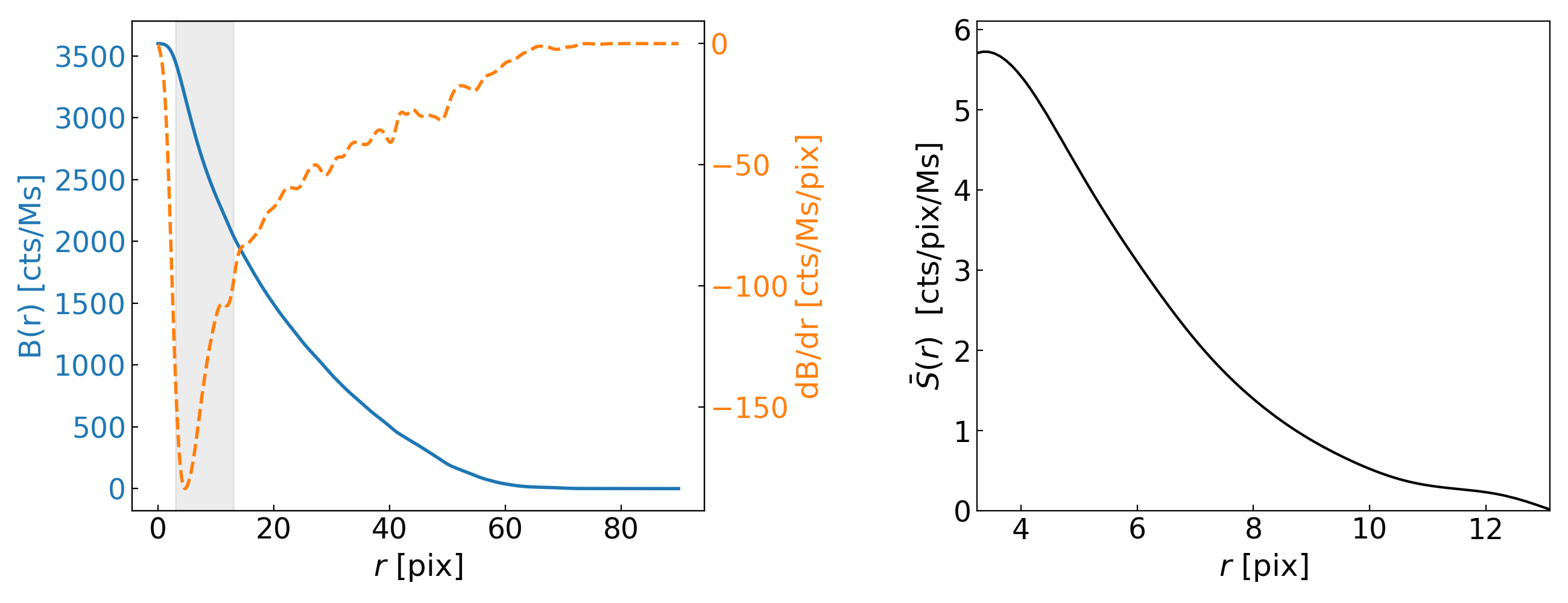}
    \caption{Left: Dependence of $B(r)$ (solid blue line, left axis), and $\frac{d\,{B}}{d\,r}$ (dashed orange line, right axis) on the SAC radius $r$. Right: $\bar S$ as a function of the SAC radius $r$ following Eq. (\ref{eq:optimal_sac}). Both panels show the $0.5-2$\,keV band results.}
    \label{fig:B_dB_Sbar}
\end{figure*}

The SNR must be an increasing function of $r$ for values of $r$ approaching the optimal exclusion radius from below.  This requires $A_F$ to be large compared to the core of the distribution of secondary particles. 
For the PN SWM we assume $A_F = 64\times 63\,\si{pixel}$.
If this condition is not met for $r \rightarrow 0$, the SNR cannot be improved by SAC. 
This assumes a uniform source distribution and ignores edge effects. 
For a uniform source distribution ($\bar S$ in Eq. \ref{eq:source} is constant) we can easily derive the optimal SAC exclusion radius $r$ based in Eq. \ref{eq:optimal_sac} if we know $B(r)$. 
We can approximate $B(r)$ based on the FWC distribution, shown in Fig. \ref{fig:sac}a), and also obtain the best achievable SNR through SAC using Eq. (\ref{eq:optimal_sac}). We derive $B(r)$ and $\frac{d\,{B}}{d\,r}$ for various energy bands (see Fig. \ref{fig:B_dB_Sbar} for the $0.5-2$\,keV band as an example), and show the optimal SNR as a function of $r$ in Fig. \ref{fig:sac_improvement} (top left). 
For each energy band a radius $r$ also corresponds to a surface brightness $\bar S$, and we indicated several values in the plot. Therefore, one can translate $r$ into $\bar S$, and in turn derive a SNR without the use of SAC for comparison. These pre-SAC values are indicated as dashed lines, and only start deviating from the SAC improved SNRs for larger $r$ and fainter sources. 
Note that the plot should not be interpreted as smaller radii resulting in a larger SNR, because each $r$ is optimized for a specific source brightness $\bar S$. 

From a user perspective Eq. \ref{eq:optimal_sac} is not intuitive, because the SAC radius $r$ is the quantity to compute. Therefore, we provide a fitting function that parameterizes the observed trend, 
\begin{equation}
    r(\bar S) = A \left( \frac{\bar S}{s_0} \right)^{-\alpha}  e^{-
    \frac{\bar S}{s_c}}
    \label{eq:s2r}
\end{equation}
We find $\alpha=0.1$ and $s_0=\SI{6.5}{cts\,pix^{-1}\,Ms^{-1}}$ to be constant for the 4 energy bands, while the amplitude $A$ and cutoff brightness $s_c$ vary: For $A$ we find $11.2$, $10.7$, $7.7$, and $\SI{8.3}{pix}$ for the $0.5-12$, $7.5-9$, $1-7$, and $0.5-2\,\si{keV}$ bands, respectively. For $s_c$ the best fit values are $15.3$, $8.2$, $10.7$, and $\SI{7.8}{cts\,pix^{-1}\,Ms^{-1}}$ for the same bands. 
The SRG/eROSITA All-Sky Survey (eRASS)\cite{Yeung2023-bc} found an average surface brightness for the diffuse sky X-ray background in the $0.5-2\,\si{keV}$ band of $\sim \SI{7e-12}{erg\,s^{-1}\,cm^{-2}\,deg^{-2}}$. For XMM PN this converts to a Case C count rate of roughly $\SI{0.1}{cts\,Ms^{-1}\,pix^{-1}}$. 
With Eq. (\ref{eq:s2r}) we can then compute the optimal SAC radius of $r=\SI{12}{pix}$.

In the top right panel of Fig. \ref{fig:sac_improvement} we show the source to background ratio, S/B, which, unlike the SNR, is independent of the exposure time. The S/B is a good indicator of background systematics and when they become dominant over the results. The SNR, on the other hand, is a statistical measure of reliability of the results. While the curves look qualitatively similar to the SNR trends, the difference from the non-SAC values is much more pronounced. We quantify the improvement (Improvement defined as $({\rm SNR_{SAC} - SNR})\cdot {\rm SNR}^{-1}$, and for S/B accordingly) of both, the SNR and S/B, in the bottom panels of Fig. \ref{fig:sac_improvement}. For fainter sources and therefore larger $r$ we find a stronger improvement of both, the SNR and S/B. While the SNR improvement is overall less significant ($<25\%$), the S/B improvement can reach 50\% and more. However, the S/B flattens at larger radii, while the SNR is increasing drastically with radius. We note that the optimal $r$ shown in Fig. \ref{fig:sac_improvement} is derived for maximizing the SNR, not the S/B. 
All four panels of Fig. \ref{fig:sac_improvement} also show the median blank sky background (as vertical bars), as estimated from the medium filter slew observations with low SP contamination. 

We verified the assumption of average slew data representing the blank sky by looking at the Case C frames of a PN SWM pointed observation of NGC 7314 (obsid 0725200101). We find that, after removing a region of 28\,arcsec radius around the bright source, the average surface brightness (after subtracting the expected Case C background events based on FWC observations) is $\SI{0.48}{cts\,Ms^{-1}\,pix^{-1}}$ in the $\SIrange{0.5}{12}{keV}$ band, which is in full agreement with the slew rates (see Fig. \ref{fig:sac_improvement}). 
The blank sky background definitely fulfills the source requirement of uniformity across the detector, and is at the faint end of all tested surface brightness values. 
We also tested filtering by event pattern, but found only a small improvement for singles exclusively. 
\begin{figure}[tbp]
    \centering
    \includegraphics[width=0.49\textwidth]{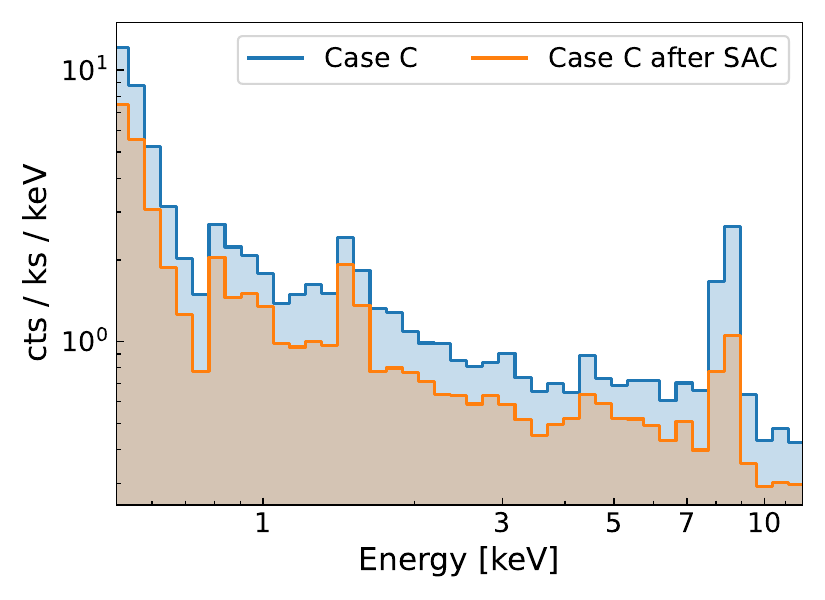}
    \caption{Case C valid event spectrum of blank sky/slew observations before and after applying the optimized SAC method. The SNR improves as expected. }
    \label{fig:sac_apply}
\end{figure}
In Figure \ref{fig:sac_apply} we illustrate the improvement of optimized SAC for the Case C spectrum of the 230\,ks medium filter slew observations. The unfiltered spectrum (blue) consists of the NXB and the average sky foreground and background emission. After applying the optimized SAC by filtering out X-ray like events based on the distance to the particle track, we find that especially the line-free energies from 1 to 7\,keV (without the Al line) has been greatly reduced (see orange curve in Fig. \ref{fig:sac_apply}). From a total of 3305 counts, 2053 are left after SAC. Based on FWC observation, we estimate a total of 2850 NXB counts, and 1584 NXB counts if SAC is applied to these observations in the same way. While the original SNR of 7.9 is expected to increase to about 9.1 (based on Fig. \ref{fig:sac_improvement}) we find that SAC indeed improves the SNR to 10.3. The source/background ratio improves from 0.16 to 0.30, where the expectation was 0.26. This demonstrates that our optimization method for SAC is viable, and can improve the signal of faint diffuse sources, in this case the sky X-ray background. 

While this experiment shows that the SNR can only be improved marginally by SAC for SWM observations, the obvious next step would be a larger detector area. Each of the 4 quadrants of the large detector of the Athena/WFI will have 64 times the area of the PN SWM region. 

\section{Summary} \label{ch:summary} 
Many characteristics of the particle induced background in X-ray observations, such as its variability, are either unknown, or have large uncertainties, which limits our ability to make optimal use of the X-ray data. 
By utilizing the PN SWM data we have significantly improved our knowledge of the NXB. The combination of a short frame time of 5.67\,ms and disabled onboard MIP rejection allowed the detected events to be classified into valid (X-ray like) and invalid (particle-like) events, based on their energy and pattern signatures. This enabled us to study the statistical properties and correlations of the NXB and particle events: 

\begin{itemize}
    \item From 626 slew observations in SWM and 519 in closed filter mode we found a distribution of particle tracks that is consistent with expectations: The spatial distribution of the tracks reflects the readout pattern, while the orientation is more uniform. However, we find that the length distribution of these particle track ``islands'' is not always consistent with a simple box detector model, where charge is equally distributed along the path through the 3D detector volume. We exclude edge effects by only selecting islands in the central part of the detector, and find that there is still an excess of shorter tracks with respect to our model, while the distribution of longer tracks is roughly consistent. The cause of these shorter particle tracks is unclear.
    \item The spectrum of valid events features several fluorescent X-ray lines, most prominently from Al, Cu and Zn. Interestingly, in frames that also contain invalid events the line flux is much higher, and the line energy of Cu and Zn is shifted to slightly lower energies, while all other fluorescent lines remain unchanged. This shift is likely due to the enhanced detection of Cu/Zn during readout, while the hole in the electronics board over the SWM aperture makes it more difficult to produce these lines during exposure to the sky. 
    \item The variability with time of the number of frames with particle tracks is well-correlated with the solar cycle. The rate of valid events shows no additional variability in FWC observations, while the rate of valid events varies by over two orders of magnitude during slew observations with the filter open. This can be attributed to soft protons. 
    \item Comparing the numbers of valid and invalid events allowed observations that are highly contaminated by soft protons to be identified among full frame slew observations (not SWM).  Using them, a model was derived for the average soft proton spectrum, consisting of a shallow powerlaw (index $-0.28$) with an exponential drop at around 9 to 10\,keV. Soft protons produce a low fraction of single events compared to normal celestial X-rays.
    \item A comparison between the detailed lightcurve of particle events in SWM and statistically superior Chandra HRC shield rates shows very good agreement. However, we detect residuals on various time scales, which require more investigation, as they might be related to changes in the GCR spectrum. No temporal offset between the detectors on the two satellites can be identified. A power spectrum analysis of the Chandra HRC shield data allows the definition of a critical timescale on the order of hours, within which the background shows negligible variability. 
    \item Frames with both particle events and X-ray like events have a special importance, since there is a high chance of the X-ray event being particle induced through secondaries. We quantify the likelihood of association between X-ray and particle events as a function of their separation. As expected, X-ray events in FWC observations are almost 100\% particle induced, while medium filter observations with low soft proton contamination have a high chance of correlation only for very small separations. A self anti-coincidence method of removing events based on the particle separation can be improved by maximizing the SNR in distinct energy bands. While for very faint sources the improvement can be substantial, the impact is greatly limited by the small aperture of the SWM. 
\end{itemize}

\section*{Disclosures}
The authors declare that there are no financial interests, commercial affiliations, or other potential conflicts of interest that could have influenced the objectivity of this research or the writing of this paper.

\section*{Code and Data Availability}
The XMM-Newton data presented in this article are publicly available through the ESAC Science Data Centre \url{https://www.cosmos.esa.int/web/xmm-newton/xsa}. Chandra HRC Anti-Coincidence shield data are available upon request. 

\section*{Acknowledgments}
We gratefully acknowledge support from NASA grant NNX17AB07G, administered by The Pennsylvania State University. 
This work was done as part of the Athena WFI Background Working Group, a consortium including MPE, INAF/IASF-Milano, IAAT, Open University, MIT, SAO, and Stanford. We would like to thank the entire Working Group for valuable discussions that contributed greatly to this paper.

This research has made use of data obtained from the Chandra Data Archive provided by the Chandra X-ray Center (CXC), and the archival data from XMM-Newton, an ESA science mission with instruments and contributions directly funded by ESA Member States and NASA. 
This research utilized the following software libraries for data analysis and presentation: XMM-Newton SAS\cite{Gabriel2004-ty}, Astropy\cite{The_Astropy_Collaboration2013-lw,The_Astropy_Collaboration2018-gx} a community-developed core Python package for Astronomy; Matplotlib\cite{Hunter2007-uf} a Python library for publication quality graphics; NumPy\cite{van-der-Walt2011-gf} a structure for eﬀicient numerical computation; SciPy\cite{Jones2001-ri} Open Source Scientific Tools for Python; and the GNU Scientific Library.

\bibliography{paperpile_remote}
\bibliographystyle{spiejour} 
%\begin{doublespacing}
%\listoffigures
%\end{doublespacing}

\vspace{2ex}\noindent\textbf{Gerrit Schellenberger} is an Astrophysicist at the Center for Astrophysics | Harvard \& Smithsonian. He received his BS and MS degrees in physics and astrophysics from  Bonn University, Germany, in 2010 and 2012, respectively, and his PhD degree in Astronomy and Astrophysics from Bonn University in 2016.  He is the author of more than 60 journal papers and is a Chandra HRC Project Scientist. His current research interests include multiwavelength observations of clusters and groups of galaxies, and systematic uncertainties in X-ray observations.

\appendix
\section{List of Small Window Mode observations}
Table \ref{tab:SWM} lists all 626 slew observation in Small Window Mode that were used for this work. Column (2) is the filter used on PN, which is either Medium (Med), Closed (FWC), or Closed with calibration source (Cal). Column (4) gives the total number of Small Window Mode frames, and Column (5) the fraction of Case A frames in per mille. 
\footnotesize
\begin{longtable}{ccccc|ccccc}
\caption{XMM-Newton PN observations in Small Window Mode during slews. \label{tab:SWM}}\\
OBSID & Filter & Date      & Total  & Case A & OBSID & Filter & Date      & Total  & Case A \\
      &        & Y/M/D     & Frames & \textperthousand &       &        & Y/M/D     & Frames & \textperthousand \\
(1)   & (2)    & (3)       & (4)    & (5)  & (1)   & (2)    & (3)       & (4)    & (5) \\
\hline
\endfirsthead
\multicolumn{10}{c}{Continuation of Table \ref{tab:SWM}}\\
(1)   & (2)    & (3)       & (4)    & (5)  & (1)   & (2)    & (3)       & (4)    & (5) \\
\hline
\endhead
9038000005 & Med & 2002/01/06 & 523468 & 6.4 & 9042500003 & Med & 2002/04/05 & 523474 & 6.3 \\
9042900002 & Med & 2002/04/13 & 523496 & 6.3 & 9044200003 & Med & 2002/05/09 & 341345 & 6.4 \\
9045000003 & Med & 2002/05/25 & 1069367 & 6.7 & 9045600004 & Med & 2002/06/06 & 329364 & 6.8 \\
9045900002 & Med & 2002/06/11 & 306525 & 6.5 & 9046100002 & Med & 2002/06/15 & 566369 & 6.7 \\
9048700002 & Med & 2002/08/06 & 262045 & 5.8 & 9050900005 & Med & 2002/09/20 & 496700 & 6.4 \\
9051000004 & Med & 2002/09/22 & 574413 & 6.4 & 9051300005 & Med & 2002/09/28 & 507731 & 6.4 \\
9053100005 & Med & 2002/11/02 & 237862 & 6.3 & 9056900004 & Med & 2003/01/17 & 865059 & 7.0 \\
9057200003 & Med & 2003/01/23 & 475199 & 6.6 & 9057200004 & Med & 2003/01/23 & 519871 & 6.3 \\
9057300003 & Med & 2003/01/24 & 310481 & 6.4 & 9057300006 & Med & 2003/01/25 & 580922 & 6.3 \\
9057500006 & Med & 2003/01/30 & 776980 & 7.0 & 9058600004 & Med & 2003/02/20 & 498185 & 6.3 \\
9059300004 & Med & 2003/03/06 & 474724 & 7.0 & 9059400002 & Med & 2003/03/08 & 334664 & 6.7 \\
9059500004 & Med & 2003/03/10 & 498657 & 6.6 & 9063300005 & Med & 2003/05/25 & 495015 & 7.0 \\
9064300004 & Med & 2003/06/14 & 551321 & 6.7 & 9066800004 & Med & 2003/08/03 & 567247 & 6.5 \\
9067100004 & Med & 2003/08/09 & 493926 & 6.5 & 9068300004 & Med & 2003/09/02 & 282502 & 6.4 \\
9069000004 & Med & 2003/09/15 & 655746 & 6.8 & 9070700004 & Med & 2003/10/19 & 250470 & 7.0 \\
9072400008 & Med & 2003/11/23 & 526419 & 5.5 & 9073500004 & Med & 2003/12/14 & 268620 & 6.1 \\
9075800004 & Med & 2004/01/29 & 280126 & 6.6 & 9076900004 & Med & 2004/02/20 & 1038827 & 7.2 \\
9077000007 & Med & 2004/02/22 & 382938 & 7.1 & 9077700004 & Med & 2004/03/07 & 208573 & 7.2 \\
9077800005 & Med & 2004/03/09 & 237174 & 7.2 & 9078000008 & Med & 2004/03/13 & 458472 & 8.4 \\
9083100003 & Med & 2004/06/22 & 682729 & 7.9 & 9083200004 & Med & 2004/06/25 & 426727 & 8.1 \\
9083700002 & Med & 2004/07/05 & 897313 & 8.2 & 9084400005 & Med & 2004/07/19 & 217153 & 8.1 \\
9084900005 & Med & 2004/07/29 & 519693 & 7.8 & 9085400004 & Med & 2004/08/08 & 480407 & 7.9 \\
9087700004 & Med & 2004/09/23 & 499699 & 7.8 & 9088200003 & Med & 2004/10/03 & 494696 & 8.8 \\
9088400004 & Med & 2004/10/06 & 881434 & 8.8 & 9088400005 & Med & 2004/10/07 & 562665 & 8.7 \\
9088800003 & Med & 2004/10/14 & 882462 & 8.7 & 9089300002 & Med & 2004/10/24 & 277918 & 8.9 \\
9089400002 & Med & 2004/10/26 & 477624 & 8.8 & 9089600003 & Med & 2004/10/30 & 571206 & 8.9 \\
9090400003 & Med & 2004/11/15 & 230488 & 7.6 & 9090800002 & Med & 2004/11/23 & 257771 & 8.9 \\
9091200002 & Med & 2004/12/01 & 1078289 & 8.8 & 9092100003 & Cal & 2004/12/19 & 683028 & 8.5 \\
9092200003 & Cal & 2004/12/21 & 721744 & 8.6 & 9093000002 & Cal & 2005/01/06 & 637169 & 7.6 \\
9093900004 & Cal & 2005/01/24 & 244076 & 7.9 & 9094100002 & Cal & 2005/01/28 & 371449 & 8.2 \\
9094100004 & Cal & 2005/01/28 & 460583 & 8.3 & 9095200004 & Cal & 2005/02/19 & 492603 & 7.8 \\
9095900002 & Med & 2005/03/04 & 961654 & 8.3 & 9095900006 & Med & 2005/03/05 & 281505 & 8.9 \\
9096000004 & Med & 2005/03/07 & 344912 & 8.6 & 9096800005 & Med & 2005/03/23 & 330852 & 7.7 \\
9097200003 & Med & 2005/03/31 & 204845 & 8.3 & 9097900002 & FWC & 2005/04/13 & 971854 & 8.6 \\
9099300005 & Med & 2005/05/12 & 662686 & 9.5 & 9100400003 & Med & 2005/06/03 & 563154 & 8.7 \\
9101400003 & Med & 2005/06/22 & 432545 & 8.9 & 9101600003 & Med & 2005/06/26 & 886969 & 8.7 \\
9101700002 & Med & 2005/06/28 & 167511 & 8.4 & 9102000010 & Med & 2005/07/05 & 313362 & 9.0 \\
9102700003 & Med & 2005/07/19 & 551682 & 7.8 & 9104200005 & Med & 2005/08/17 & 325331 & 8.3 \\
9104700006 & Med & 2005/08/28 & 1010313 & 8.5 & 9106100004 & Med & 2005/09/24 & 342233 & 7.8 \\
9106300005 & Med & 2005/09/29 & 178251 & 8.3 & 9106500002 & Med & 2005/10/02 & 526588 & 8.4 \\
9107300002 & Med & 2005/10/19 & 715510 & 9.1 & 9108200004 & Med & 2005/11/05 & 160843 & 8.9 \\
9109500002 & Med & 2005/12/01 & 1000085 & 9.3 & 9112900003 & Med & 2006/02/07 & 374145 & 10.0 \\
9114200002 & Med & 2006/03/05 & 203674 & 10.3 & 9114900002 & Med & 2006/03/19 & 231332 & 10.4 \\
9115300002 & Med & 2006/03/27 & 311262 & 10.2 & 9115800002 & Med & 2006/04/06 & 477591 & 10.3 \\
9116500004 & Med & 2006/04/20 & 1172741 & 10.5 & 9116800003 & Med & 2006/04/26 & 937027 & 10.6 \\
9119300003 & Med & 2006/06/14 & 953546 & 10.7 & 9120200004 & Med & 2006/07/03 & 302797 & 11.0 \\
9122300002 & Med & 2006/08/13 & 559123 & 10.8 & 9122400002 & Med & 2006/08/15 & 432807 & 10.8 \\
9122500003 & Med & 2006/08/17 & 221949 & 10.7 & 9122700002 & Med & 2006/08/21 & 522650 & 10.1 \\
9124600002 & Med & 2006/09/28 & 266130 & 10.8 & 9124700002 & Med & 2006/09/30 & 761052 & 10.6 \\
9124700003 & Med & 2006/09/30 & 1176258 & 10.6 & 9125100003 & Med & 2006/10/08 & 620406 & 10.9 \\
9125400002 & Med & 2006/10/14 & 252815 & 10.8 & 9125700003 & Med & 2006/10/20 & 825923 & 11.1 \\
9125900002 & Med & 2006/10/24 & 1116477 & 11.0 & 9126300004 & Med & 2006/11/01 & 186605 & 11.1 \\
9126300005 & Med & 2006/11/02 & 211586 & 11.3 & 9126400002 & Med & 2006/11/03 & 298936 & 11.1 \\
9126500002 & Med & 2006/11/05 & 369890 & 11.2 & 9126600002 & Med & 2006/11/08 & 680432 & 10.9 \\
9130000002 & Med & 2007/01/13 & 1141952 & 11.2 & 9131100004 & Med & 2007/02/05 & 744510 & 11.1 \\
9131300004 & Med & 2007/02/09 & 1122224 & 10.9 & 9132900004 & Med & 2007/03/13 & 722415 & 9.9 \\
9133000003 & Med & 2007/03/15 & 177891 & 11.0 & 9133200004 & Med & 2007/03/19 & 361499 & 11.5 \\
9133300003 & Med & 2007/03/21 & 204769 & 11.4 & 9134300002 & Med & 2007/04/10 & 411539 & 11.6 \\
9134700002 & Med & 2007/04/18 & 693924 & 11.6 & 9134900002 & Med & 2007/04/22 & 414767 & 11.7 \\
9136000003 & FWC & 2007/05/13 & 690115 & 12.0 & 9136100002 & FWC & 2007/05/16 & 418639 & 12.0 \\
9136200004 & FWC & 2007/05/18 & 1132077 & 11.9 & 9136500003 & FWC & 2007/05/23 & 717519 & 11.5 \\
9137500005 & FWC & 2007/06/12 & 995629 & 11.9 & 9138800003 & FWC & 2007/07/08 & 395751 & 12.2 \\
9138900004 & FWC & 2007/07/11 & 778930 & 11.9 & 9139200003 & FWC & 2007/07/17 & 504354 & 11.9 \\
9139400002 & FWC & 2007/07/20 & 201915 & 11.8 & 9139500004 & FWC & 2007/07/23 & 759509 & 11.9 \\
9139700002 & FWC & 2007/07/26 & 383898 & 12.3 & 9140100004 & FWC & 2007/08/03 & 241881 & 11.9 \\
9141000003 & FWC & 2007/08/21 & 719021 & 12.1 & 9142800004 & FWC & 2007/09/26 & 1005948 & 12.1 \\
9143300002 & FWC & 2007/10/06 & 654472 & 12.2 & 9144300004 & FWC & 2007/10/26 & 1087657 & 12.0 \\
9144500003 & FWC & 2007/10/30 & 522940 & 12.0 & 9144700003 & FWC & 2007/11/03 & 175325 & 11.7 \\
9144900005 & FWC & 2007/11/08 & 234024 & 13.0 & 9145700003 & FWC & 2007/11/23 & 917227 & 11.7 \\
9146300006 & FWC & 2007/12/05 & 587071 & 12.4 & 9147500002 & FWC & 2007/12/28 & 330882 & 12.6 \\
9147900002 & FWC & 2008/01/05 & 327374 & 12.5 & 9148000004 & FWC & 2008/01/08 & 225990 & 11.7 \\
9148400003 & FWC & 2008/01/16 & 440017 & 11.9 & 9149500002 & FWC & 2008/02/07 & 838018 & 12.1 \\
9151000002 & FWC & 2008/03/08 & 1175237 & 11.8 & 9151000003 & FWC & 2008/03/08 & 632727 & 12.1 \\
9151300002 & FWC & 2008/03/14 & 922996 & 11.7 & 9151600004 & FWC & 2008/03/20 & 349685 & 11.9 \\
9151700002 & FWC & 2008/03/22 & 168322 & 12.4 & 9152300002 & FWC & 2008/04/02 & 377645 & 11.9 \\
9152400002 & FWC & 2008/04/04 & 585622 & 11.9 & 9152700003 & FWC & 2008/04/11 & 667737 & 11.8 \\
9152900002 & FWC & 2008/04/14 & 624744 & 11.9 & 9153000003 & FWC & 2008/04/16 & 735807 & 11.9 \\
9153100004 & FWC & 2008/04/19 & 1007745 & 12.0 & 9153200003 & FWC & 2008/04/21 & 936524 & 12.1 \\
9153300002 & FWC & 2008/04/22 & 827870 & 12.1 & 9153400002 & FWC & 2008/04/25 & 652006 & 11.8 \\
9153400004 & FWC & 2008/04/25 & 868326 & 11.7 & 9153600002 & FWC & 2008/04/28 & 964646 & 12.1 \\
9153600003 & FWC & 2008/04/29 & 1090252 & 12.1 & 9153900002 & FWC & 2008/05/05 & 184263 & 11.9 \\
9154200004 & FWC & 2008/05/11 & 1103387 & 12.1 & 9154300003 & FWC & 2008/05/13 & 510530 & 12.0 \\
9154400005 & FWC & 2008/05/15 & 536232 & 12.1 & 9154600005 & FWC & 2008/05/19 & 773230 & 12.2 \\
9156800003 & FWC & 2008/07/02 & 1027722 & 12.3 & 9158100002 & FWC & 2008/07/28 & 822192 & 12.2 \\
9158900004 & FWC & 2008/08/12 & 589018 & 12.2 & 9160000002 & FWC & 2008/09/03 & 578524 & 12.9 \\
9160700004 & FWC & 2008/09/17 & 253916 & 12.7 & 9160800004 & FWC & 2008/09/20 & 1025948 & 12.9 \\
9160900002 & FWC & 2008/09/21 & 519505 & 12.8 & 9161000002 & FWC & 2008/09/23 & 167731 & 12.7 \\
9161300002 & FWC & 2008/09/29 & 189574 & 12.6 & 9161500004 & FWC & 2008/10/03 & 358285 & 12.6 \\
9161600002 & FWC & 2008/10/05 & 1062545 & 12.6 & 9161900002 & FWC & 2008/10/11 & 268946 & 13.2 \\
9162100003 & FWC & 2008/10/15 & 617166 & 12.8 & 9163100002 & FWC & 2008/11/04 & 341052 & 12.6 \\
9164900002 & FWC & 2008/12/10 & 450612 & 12.9 & 9164900003 & FWC & 2008/12/10 & 536479 & 12.8 \\
9165500004 & FWC & 2008/12/22 & 408866 & 12.8 & 9166200003 & FWC & 2009/01/05 & 175344 & 12.7 \\
9168100003 & FWC & 2009/02/12 & 260532 & 13.6 & 9169500002 & FWC & 2009/03/12 & 245522 & 13.2 \\
9169600003 & FWC & 2009/03/14 & 273413 & 13.1 & 9169700004 & FWC & 2009/03/16 & 591747 & 13.2 \\
9169800002 & FWC & 2009/03/17 & 725286 & 13.4 & 9169900004 & FWC & 2009/03/20 & 804165 & 13.4 \\
9170200002 & FWC & 2009/03/25 & 362767 & 13.3 & 9170300002 & FWC & 2009/03/28 & 816309 & 13.4 \\
9170500003 & FWC & 2009/03/31 & 700702 & 13.6 & 9171000002 & FWC & 2009/04/10 & 926909 & 13.5 \\
9171000003 & FWC & 2009/04/11 & 450835 & 13.3 & 9171100004 & FWC & 2009/04/13 & 225876 & 13.2 \\
9171600003 & FWC & 2009/04/23 & 756501 & 13.6 & 9172300002 & FWC & 2009/05/06 & 720995 & 13.4 \\
9173400002 & FWC & 2009/05/29 & 1180658 & 13.7 & 9175700002 & FWC & 2009/07/14 & 826809 & 14.0 \\
9176600004 & FWC & 2009/08/01 & 346700 & 13.6 & 9176700003 & FWC & 2009/08/03 & 472923 & 14.1 \\
9176800004 & FWC & 2009/08/05 & 300521 & 13.8 & 9176900004 & FWC & 2009/08/07 & 285291 & 13.4 \\
9177600004 & FWC & 2009/08/20 & 603084 & 13.5 & 9178100003 & FWC & 2009/08/31 & 309352 & 13.6 \\
9179300002 & FWC & 2009/09/23 & 591640 & 13.6 & 9180400003 & FWC & 2009/10/15 & 454489 & 13.8 \\
9180700003 & FWC & 2009/10/21 & 744110 & 14.0 & 9181300003 & FWC & 2009/11/02 & 951097 & 13.9 \\
9181400002 & FWC & 2009/11/04 & 209205 & 14.0 & 9181500003 & FWC & 2009/11/07 & 832036 & 13.9 \\
9181700003 & FWC & 2009/11/11 & 415449 & 14.0 & 9181900003 & FWC & 2009/11/15 & 740379 & 14.0 \\
9182100003 & FWC & 2009/11/18 & 928535 & 14.2 & 9182200003 & FWC & 2009/11/21 & 848690 & 14.0 \\
9182500003 & FWC & 2009/11/26 & 534366 & 14.0 & 9185700003 & FWC & 2010/01/29 & 267074 & 13.6 \\
9187200003 & FWC & 2010/02/28 & 405573 & 13.5 & 9187300003 & FWC & 2010/03/02 & 591189 & 13.4 \\
9187400002 & FWC & 2010/03/04 & 751984 & 13.1 & 9187400003 & FWC & 2010/03/04 & 443820 & 13.2 \\
9188300003 & FWC & 2010/03/22 & 192226 & 12.7 & 9189200004 & FWC & 2010/04/09 & 668569 & 12.2 \\
9190100002 & FWC & 2010/04/27 & 882667 & 12.5 & 9190400003 & FWC & 2010/05/03 & 244769 & 12.5 \\
9190600003 & FWC & 2010/05/07 & 409138 & 12.2 & 9191000002 & FWC & 2010/05/15 & 1134923 & 12.7 \\
9191100005 & FWC & 2010/05/17 & 549655 & 12.8 & 9191300004 & FWC & 2010/05/21 & 877200 & 12.3 \\
9191600003 & FWC & 2010/05/27 & 498337 & 12.4 & 9191700004 & FWC & 2010/05/29 & 1010112 & 11.9 \\
9191800002 & FWC & 2010/05/31 & 1010659 & 12.4 & 9192100003 & FWC & 2010/06/06 & 397557 & 12.0 \\
9193100002 & FWC & 2010/06/26 & 377031 & 12.0 & 9193200002 & FWC & 2010/06/28 & 545220 & 12.3 \\
9194500007 & FWC & 2010/07/24 & 380832 & 12.4 & 9194800004 & FWC & 2010/07/29 & 625569 & 12.0 \\
9195000003 & FWC & 2010/08/02 & 658117 & 12.2 & 9196600002 & FWC & 2010/09/04 & 913746 & 12.2 \\
9196900002 & FWC & 2010/09/09 & 220959 & 12.6 & 9197000002 & FWC & 2010/09/11 & 929152 & 12.1 \\
9197500003 & FWC & 2010/09/21 & 399390 & 11.7 & 9198100002 & FWC & 2010/10/03 & 390516 & 12.0 \\
9198300002 & FWC & 2010/10/07 & 716620 & 11.8 & 9198400003 & FWC & 2010/10/10 & 443393 & 12.2 \\
9198700006 & FWC & 2010/10/16 & 192493 & 12.1 & 9198900002 & FWC & 2010/10/19 & 320276 & 12.0 \\
9198900004 & FWC & 2010/10/20 & 780709 & 11.9 & 9199200004 & FWC & 2010/10/25 & 185636 & 11.8 \\
9199500004 & FWC & 2010/10/31 & 336063 & 11.1 & 9200100005 & FWC & 2010/11/12 & 563142 & 11.8 \\
9200200002 & FWC & 2010/11/14 & 568866 & 11.6 & 9200400003 & FWC & 2010/11/18 & 373046 & 11.5 \\
9200900003 & FWC & 2010/11/28 & 307234 & 11.6 & 9201300003 & FWC & 2010/12/07 & 730248 & 11.6 \\
9201400003 & FWC & 2010/12/08 & 292366 & 11.8 & 9201500003 & FWC & 2010/12/10 & 1144808 & 11.7 \\
9202100003 & FWC & 2010/12/22 & 330234 & 11.6 & 9202900002 & FWC & 2011/01/06 & 741634 & 11.6 \\
9204700002 & FWC & 2011/02/12 & 282536 & 11.7 & 9205700003 & FWC & 2011/03/04 & 1121528 & 11.4 \\
9207100003 & FWC & 2011/03/31 & 463037 & 10.8 & 9207600004 & FWC & 2011/04/11 & 299694 & 10.8 \\
9207700003 & FWC & 2011/04/13 & 612780 & 10.2 & 9208100004 & FWC & 2011/04/21 & 1051383 & 10.5 \\
9208400003 & FWC & 2011/04/27 & 942753 & 10.8 & 9209500004 & FWC & 2011/05/19 & 561570 & 10.9 \\
9209600002 & FWC & 2011/05/20 & 1155921 & 10.9 & 9209800002 & FWC & 2011/05/24 & 957568 & 11.0 \\
9210100003 & FWC & 2011/05/31 & 744293 & 10.4 & 9210700004 & FWC & 2011/06/12 & 696953 & 10.5 \\
9211600002 & FWC & 2011/06/29 & 1047641 & 9.8 & 9211700002 & FWC & 2011/07/01 & 376951 & 10.0 \\
9214900004 & FWC & 2011/09/04 & 249888 & 10.3 & 9218200004 & FWC & 2011/11/08 & 454187 & 10.3 \\
9218300002 & FWC & 2011/11/10 & 435766 & 10.2 & 9223300002 & FWC & 2012/02/18 & 935851 & 9.9 \\
9225900003 & FWC & 2012/04/10 & 560296 & 9.7 & 9226100002 & FWC & 2012/04/14 & 1035930 & 9.9 \\
9226400004 & FWC & 2012/04/20 & 464484 & 10.3 & 9227500006 & FWC & 2012/05/12 & 611512 & 9.7 \\
9227600002 & FWC & 2012/05/13 & 394875 & 9.6 & 9229000002 & FWC & 2012/06/10 & 324826 & 9.0 \\
9229600003 & FWC & 2012/06/22 & 323646 & 9.3 & 9229700003 & FWC & 2012/06/25 & 267659 & 9.8 \\
9229900003 & FWC & 2012/06/28 & 265961 & 9.7 & 9230100004 & FWC & 2012/07/03 & 260672 & 9.5 \\
9231800002 & FWC & 2012/08/05 & 673050 & 8.1 & 9232100004 & FWC & 2012/08/11 & 1199939 & 8.3 \\
9233000003 & FWC & 2012/08/29 & 403990 & 8.7 & 9233900005 & FWC & 2012/09/16 & 782485 & 8.7 \\
9236300002 & FWC & 2012/11/03 & 669150 & 8.7 & 9236600002 & FWC & 2012/11/09 & 1042098 & 9.0 \\
9236700002 & FWC & 2012/11/10 & 1144366 & 8.9 & 9236900002 & FWC & 2012/11/14 & 865821 & 8.5 \\
9238200002 & FWC & 2012/12/10 & 1038779 & 8.7 & 9238700004 & FWC & 2012/12/21 & 766697 & 8.6 \\
9239400003 & FWC & 2013/01/04 & 678821 & 9.0 & 9240900002 & FWC & 2013/02/03 & 595050 & 9.3 \\
9241200002 & FWC & 2013/02/08 & 371796 & 8.9 & 9241500002 & FWC & 2013/02/14 & 944903 & 9.3 \\
9241600002 & FWC & 2013/02/16 & 196163 & 9.6 & 9242200003 & FWC & 2013/03/01 & 359328 & 9.1 \\
9242700002 & FWC & 2013/03/11 & 401728 & 9.5 & 9245700004 & FWC & 2013/05/10 & 405283 & 8.9 \\
9247900002 & FWC & 2013/06/22 & 182171 & 7.8 & 9248700002 & FWC & 2013/07/08 & 216005 & 8.1 \\
9248900002 & FWC & 2013/07/12 & 425182 & 7.5 & 9248900003 & FWC & 2013/07/12 & 640578 & 7.6 \\
9249100002 & FWC & 2013/07/16 & 1189396 & 7.6 & 9249300002 & FWC & 2013/07/20 & 1106856 & 8.1 \\
9249400002 & FWC & 2013/07/22 & 603515 & 8.2 & 9249500003 & FWC & 2013/07/25 & 1073203 & 8.2 \\
9249600002 & FWC & 2013/07/26 & 614271 & 8.2 & 9249700002 & FWC & 2013/07/28 & 615427 & 8.1 \\
9249800002 & FWC & 2013/07/30 & 302452 & 7.8 & 9249900002 & FWC & 2013/08/01 & 708997 & 7.9 \\
9254500004 & FWC & 2013/11/01 & 723226 & 7.9 & 9254600004 & FWC & 2013/11/03 & 830218 & 8.0 \\
9256500002 & FWC & 2013/12/11 & 1028960 & 7.6 & 9256600002 & FWC & 2013/12/13 & 764499 & 7.8 \\
9257300002 & FWC & 2013/12/26 & 274812 & 7.7 & 9258700002 & FWC & 2014/01/24 & 751265 & 7.8 \\
9258800002 & FWC & 2014/01/26 & 557815 & 7.7 & 9259300002 & FWC & 2014/02/05 & 1018256 & 7.6 \\
9261200003 & FWC & 2014/03/15 & 950946 & 7.2 & 9261600002 & Med & 2014/03/22 & 526843 & 7.7 \\
9261700002 & Med & 2014/03/24 & 918132 & 7.7 & 9262500003 & FWC & 2014/04/10 & 1046692 & 7.7 \\
9263300002 & FWC & 2014/04/25 & 271099 & 7.5 & 9264200002 & FWC & 2014/05/13 & 219010 & 7.9 \\
9264400003 & FWC & 2014/05/17 & 737024 & 8.1 & 9265000003 & FWC & 2014/05/29 & 908458 & 7.8 \\
9265500003 & FWC & 2014/06/09 & 396031 & 7.8 & 9266200002 & FWC & 2014/06/22 & 276761 & 7.4 \\
9266700002 & FWC & 2014/07/02 & 531491 & 7.7 & 9267800004 & FWC & 2014/07/24 & 800161 & 8.1 \\
9268600003 & FWC & 2014/08/09 & 1139448 & 7.8 & 9268900002 & FWC & 2014/08/15 & 993992 & 8.1 \\
9269100003 & FWC & 2014/08/19 & 219321 & 8.4 & 9269300002 & FWC & 2014/08/23 & 769649 & 8.3 \\
9270200002 & FWC & 2014/09/09 & 508512 & 8.1 & 9272200003 & FWC & 2014/10/20 & 689469 & 9.0 \\
9272300003 & FWC & 2014/10/22 & 316186 & 8.9 & 9272400004 & FWC & 2014/10/24 & 1093791 & 8.3 \\
9273200003 & FWC & 2014/11/09 & 842060 & 8.2 & 9273400004 & FWC & 2014/11/13 & 1160736 & 8.2 \\
9274300003 & FWC & 2014/12/01 & 1002277 & 8.4 & 9274700002 & FWC & 2014/12/09 & 1008580 & 8.0 \\
9275600002 & FWC & 2014/12/27 & 1109659 & 7.6 & 9275900004 & FWC & 2015/01/02 & 1066102 & 7.7 \\
9276000005 & FWC & 2015/01/04 & 1046142 & 7.9 & 9276100002 & FWC & 2015/01/05 & 167170 & 7.9 \\
9276400002 & FWC & 2015/01/11 & 492080 & 8.5 & 9276600002 & FWC & 2015/01/16 & 668181 & 8.8 \\
9276600003 & FWC & 2015/01/16 & 611061 & 8.5 & 9276700003 & FWC & 2015/01/18 & 828185 & 8.6 \\
9278000004 & FWC & 2015/02/13 & 607387 & 8.4 & 9278900002 & FWC & 2015/03/03 & 703201 & 7.8 \\
9280600003 & FWC & 2015/04/06 & 349926 & 7.9 & 9281000002 & FWC & 2015/04/13 & 552695 & 7.5 \\
9281200003 & FWC & 2015/04/18 & 637549 & 7.5 & 9281300003 & FWC & 2015/04/20 & 384110 & 7.6 \\
9285000003 & FWC & 2015/07/02 & 948373 & 8.6 & 9285400002 & FWC & 2015/07/10 & 176059 & 9.0 \\
9285400003 & FWC & 2015/07/10 & 559730 & 8.5 & 9285600002 & FWC & 2015/07/14 & 173366 & 8.4 \\
9285700003 & FWC & 2015/07/16 & 669302 & 8.5 & 9288200003 & FWC & 2015/09/05 & 191254 & 9.0 \\
9289500004 & FWC & 2015/09/30 & 305504 & 9.0 & 9289800002 & FWC & 2015/10/06 & 200066 & 8.8 \\
9290800002 & FWC & 2015/10/26 & 608146 & 8.8 & 9291000004 & FWC & 2015/10/30 & 920331 & 9.3 \\
9291100003 & FWC & 2015/11/01 & 529684 & 9.3 & 9291500002 & FWC & 2015/11/08 & 181708 & 8.7 \\
9291600003 & FWC & 2015/11/11 & 491658 & 8.5 & 9291600004 & FWC & 2015/11/11 & 530926 & 8.6 \\
9291700002 & FWC & 2015/11/13 & 272614 & 8.6 & 9291800002 & FWC & 2015/11/15 & 867688 & 8.9 \\
9291900002 & FWC & 2015/11/17 & 470589 & 9.0 & 9292200002 & FWC & 2015/11/23 & 921948 & 9.2 \\
9292300003 & FWC & 2015/11/25 & 579247 & 9.7 & 9292400005 & FWC & 2015/11/27 & 624384 & 9.5 \\
9293100002 & FWC & 2015/12/11 & 1118775 & 8.9 & 9293400002 & FWC & 2015/12/16 & 196288 & 9.2 \\
9293500002 & FWC & 2015/12/18 & 213549 & 9.4 & 9293700002 & FWC & 2015/12/22 & 844385 & 9.5 \\
9294700014 & FWC & 2016/01/12 & 319966 & 9.4 & 9294800004 & FWC & 2016/01/14 & 282728 & 9.8 \\
9294900005 & FWC & 2016/01/16 & 710342 & 9.7 & 9295700004 & Med & 2016/02/01 & 305377 & 10.2 \\
9296400004 & Med & 2016/02/15 & 191602 & 10.2 & 9299500002 & Med & 2016/04/16 & 297925 & 10.1 \\
9299500003 & Med & 2016/04/16 & 625763 & 10.3 & 9299800004 & Med & 2016/04/23 & 164251 & 10.2 \\
9300500002 & Med & 2016/05/06 & 236379 & 10.1 & 9300500004 & Med & 2016/05/06 & 1016731 & 10.7 \\
9300900004 & Med & 2016/05/15 & 504497 & 10.8 & 9301000003 & Med & 2016/05/16 & 418266 & 10.8 \\
9301200005 & Med & 2016/05/21 & 1170913 & 10.2 & 9301400002 & Med & 2016/05/24 & 197225 & 10.4 \\
9301500005 & Med & 2016/05/27 & 348926 & 10.8 & 9302200006 & Med & 2016/06/10 & 676012 & 11.1 \\
9302400003 & Med & 2016/06/13 & 578355 & 10.9 & 9302900003 & Med & 2016/06/24 & 173836 & 10.4 \\
9303100002 & Med & 2016/06/27 & 320084 & 10.5 & 9303200005 & Med & 2016/06/29 & 718162 & 10.9 \\
9304200004 & Med & 2016/07/19 & 985842 & 10.4 & 9304200005 & Med & 2016/07/19 & 244540 & 11.0 \\
9305600003 & FWC & 2016/08/16 & 767599 & 10.8 & 9305600004 & FWC & 2016/08/16 & 807778 & 11.0 \\
9305700003 & FWC & 2016/08/18 & 943507 & 10.7 & 9305700005 & FWC & 2016/08/18 & 352480 & 11.0 \\
9305800002 & FWC & 2016/08/20 & 843569 & 10.9 & 9306300003 & FWC & 2016/08/30 & 405606 & 11.8 \\
9306400004 & FWC & 2016/09/01 & 1102454 & 11.2 & 9307500002 & FWC & 2016/09/22 & 438809 & 11.6 \\
9307800002 & FWC & 2016/09/28 & 285355 & 11.3 & 9307900004 & FWC & 2016/10/01 & 546743 & 10.8 \\
9307900005 & FWC & 2016/10/01 & 1142451 & 11.1 & 9308100004 & FWC & 2016/10/05 & 641869 & 11.4 \\
9308100005 & FWC & 2016/10/05 & 442307 & 11.3 & 9308700003 & FWC & 2016/10/17 & 557726 & 11.3 \\
9309200003 & FWC & 2016/10/27 & 446345 & 11.4 & 9309500002 & FWC & 2016/11/01 & 518427 & 11.5 \\
9309900002 & FWC & 2016/11/10 & 804167 & 11.8 & 9310200004 & FWC & 2016/11/16 & 830683 & 12.0 \\
9311100002 & FWC & 2016/12/03 & 274180 & 11.9 & 9311100005 & FWC & 2016/12/04 & 883407 & 12.2 \\
9311600004 & FWC & 2016/12/14 & 427108 & 12.4 & 9312000002 & FWC & 2016/12/22 & 164444 & 11.8 \\
9312000003 & FWC & 2016/12/22 & 542822 & 11.9 & 9312000004 & FWC & 2016/12/22 & 358101 & 11.8 \\
9313500002 & FWC & 2017/01/20 & 800550 & 12.3 & 9315100002 & FWC & 2017/02/22 & 267558 & 12.7 \\
9316200002 & FWC & 2017/03/16 & 857240 & 12.9 & 9316200003 & FWC & 2017/03/16 & 273717 & 12.5 \\
9317200002 & FWC & 2017/04/04 & 271624 & 11.9 & 9319100004 & FWC & 2017/05/13 & 725326 & 12.9 \\
9321200004 & FWC & 2017/06/23 & 654216 & 13.1 & 9321700003 & FWC & 2017/07/03 & 348437 & 12.2 \\
9322400005 & FWC & 2017/07/18 & 768957 & 12.0 & 9323600003 & FWC & 2017/08/10 & 402662 & 12.3 \\
9324200003 & FWC & 2017/08/22 & 354198 & 11.4 & 9324400004 & FWC & 2017/08/26 & 1115586 & 11.8 \\
9325500002 & FWC & 2017/09/17 & 143721 & 11.2 & 9327000002 & FWC & 2017/10/17 & 796044 & 11.8 \\
9327600004 & FWC & 2017/10/29 & 179335 & 12.0 & 9328000003 & FWC & 2017/11/06 & 896198 & 12.2 \\
9328400002 & FWC & 2017/11/13 & 309746 & 12.1 & 9328700006 & FWC & 2017/11/20 & 820495 & 12.7 \\
9329500002 & FWC & 2017/12/05 & 145165 & 12.6 & 9330400003 & FWC & 2017/12/24 & 140402 & 12.8 \\
9331600004 & FWC & 2018/01/16 & 342229 & 13.1 & 9333900003 & FWC & 2018/03/03 & 241550 & 13.2 \\
9334400004 & FWC & 2018/03/14 & 1109132 & 13.4 & 9336200003 & FWC & 2018/04/18 & 250776 & 13.3 \\
9336800004 & Med & 2018/05/01 & 138831 & 13.1 & 9337300006 & FWC & 2018/05/10 & 290849 & 13.0 \\
9337400003 & FWC & 2018/05/13 & 310356 & 13.6 & 9337500004 & FWC & 2018/05/15 & 249747 & 13.5 \\
9337700003 & FWC & 2018/05/18 & 172260 & 13.1 & 9338000002 & FWC & 2018/05/24 & 396609 & 12.9 \\
9338400002 & FWC & 2018/06/01 & 713152 & 13.7 & 9339200003 & FWC & 2018/06/17 & 585428 & 13.9 \\
9339700003 & FWC & 2018/06/28 & 669044 & 12.7 & 9341100003 & FWC & 2018/07/26 & 611272 & 13.3 \\
9342500003 & FWC & 2018/08/22 & 390765 & 13.6 & 9342700002 & FWC & 2018/08/26 & 319624 & 13.6 \\
9342900002 & FWC & 2018/08/30 & 739933 & 13.4 & 9345300003 & FWC & 2018/10/17 & 315345 & 13.6 \\
9345900003 & FWC & 2018/10/29 & 608707 & 13.8 & 9346400002 & FWC & 2018/11/08 & 742179 & 13.5 \\
9346500002 & FWC & 2018/11/10 & 1043936 & 13.3 & 9346700003 & FWC & 2018/11/14 & 336528 & 13.2 \\
9347100007 & FWC & 2018/11/22 & 958020 & 13.8 & 9347100008 & FWC & 2018/11/22 & 298778 & 14.0 \\
9347300003 & FWC & 2018/11/25 & 319035 & 13.6 & 9348600002 & FWC & 2018/12/21 & 716599 & 13.8 \\
9352800003 & FWC & 2019/03/15 & 240520 & 13.8 & 9354500004 & FWC & 2019/04/18 & 652828 & 14.0 \\
9354900002 & FWC & 2019/04/26 & 197567 & 13.9 & 9355400007 & FWC & 2019/05/06 & 428732 & 14.0 \\
9355500002 & FWC & 2019/05/09 & 577381 & 13.6 & 9355600002 & FWC & 2019/05/11 & 577058 & 13.2 \\
9356200004 & FWC & 2019/05/22 & 421150 & 13.6 & 9358300002 & FWC & 2019/07/03 & 1026272 & 13.9 \\
9358900003 & FWC & 2019/07/15 & 249014 & 13.8 & 9359200002 & FWC & 2019/07/21 & 241940 & 14.4 \\
9359200004 & FWC & 2019/07/21 & 522385 & 14.2 & 9359600003 & FWC & 2019/07/30 & 998555 & 14.0 \\
9359800003 & FWC & 2019/08/02 & 1149916 & 14.1 & 9360200002 & FWC & 2019/08/10 & 540230 & 13.8 \\
9360600002 & FWC & 2019/08/18 & 580090 & 14.1 & 9361100002 & FWC & 2019/08/28 & 437059 & 14.0 \\
9362400002 & FWC & 2019/09/23 & 631360 & 14.4 & 9362600002 & FWC & 2019/09/26 & 174873 & 14.1 \\
9362700005 & FWC & 2019/09/29 & 380936 & 14.0 & 9363300002 & FWC & 2019/10/10 & 1008347 & 14.2 \\
9363500002 & FWC & 2019/10/15 & 554776 & 14.3 & 9364100003 & FWC & 2019/10/27 & 957774 & 13.9 \\
9364200003 & FWC & 2019/10/29 & 685418 & 14.1 & 9364200004 & FWC & 2019/10/29 & 269886 & 14.1 \\
9364300004 & FWC & 2019/10/31 & 614396 & 14.3 & 9364400003 & FWC & 2019/11/02 & 594039 & 13.8 \\
9364600002 & FWC & 2019/11/05 & 605596 & 14.1 & 9364600003 & FWC & 2019/11/06 & 412003 & 14.3 \\
9364700002 & FWC & 2019/11/07 & 1201808 & 14.3 & 9364800004 & FWC & 2019/11/10 & 428537 & 14.5 \\
9365200002 & FWC & 2019/11/17 & 698482 & 14.2 & 9365200004 & FWC & 2019/11/18 & 393143 & 14.4 \\
9365300002 & FWC & 2019/11/19 & 805944 & 14.3 & 9366100004 & FWC & 2019/12/06 & 203332 & 14.4 \\
9366300002 & FWC & 2019/12/09 & 685067 & 14.3 & 9366400003 & FWC & 2019/12/12 & 182759 & 14.2 \\
9368200002 & FWC & 2020/01/16 & 565202 & 14.2 & 9368300003 & FWC & 2020/01/19 & 346330 & 14.4 \\
9369800002 & FWC & 2020/02/17 & 257636 & 14.6 & 9371100002 & FWC & 2020/03/15 & 336133 & 14.4 \\
9372000002 & FWC & 2020/04/01 & 282765 & 14.4 & 9372300003 & FWC & 2020/04/07 & 589725 & 14.8 \\
9372500002 & FWC & 2020/04/11 & 231638 & 14.5 & 9373200002 & FWC & 2020/04/25 & 176451 & 14.8 \\
9373600003 & FWC & 2020/05/03 & 533969 & 14.1 & 9374300002 & FWC & 2020/05/18 & 808119 & 14.5 \\
9375000002 & FWC & 2020/06/01 & 837827 & 14.5 & 9376100003 & FWC & 2020/06/22 & 546639 & 14.8 \\
9377600003 & FWC & 2020/07/22 & 492169 & 14.9 & 9380900002 & FWC & 2020/09/25 & 1045963 & 13.9 \\
9381100002 & FWC & 2020/09/30 & 232651 & 13.8 & 9382600002 & FWC & 2020/10/29 & 1121726 & 14.1 \\
9382600004 & FWC & 2020/10/30 & 333598 & 14.5 & 9383600002 & FWC & 2020/11/18 & 853611 & 14.4 \\
9383900002 & FWC & 2020/11/24 & 260636 & 14.0 & 9384200002 & FWC & 2020/12/01 & 792131 & 15.1 \\
9384300002 & FWC & 2020/12/03 & 983104 & 14.6 & 9385000002 & FWC & 2020/12/16 & 183055 & 14.2 \\
9385200002 & FWC & 2020/12/20 & 341423 & 13.7 & 9385300003 & FWC & 2020/12/22 & 1092919 & 13.8 \\
9385400003 & FWC & 2020/12/25 & 283229 & 14.0 & 9385900003 & FWC & 2021/01/03 & 192986 & 13.9 \\
9386100003 & FWC & 2021/01/08 & 634424 & 13.9 & 9386300004 & FWC & 2021/01/12 & 1017226 & 13.4 \\
9389400003 & FWC & 2021/03/15 & 836867 & 13.7 & 9389500003 & FWC & 2021/03/16 & 215495 & 13.8 \\
9389600003 & FWC & 2021/03/18 & 191737 & 13.8 & 9389900003 & FWC & 2021/03/24 & 190492 & 13.9 \\
9390000004 & FWC & 2021/03/26 & 177079 & 13.2 & 9390000006 & FWC & 2021/03/27 & 169555 & 13.7 \\
9390300003 & FWC & 2021/04/02 & 180241 & 14.2 & 9391300003 & FWC & 2021/04/21 & 212539 & 13.3 \\
9391400004 & FWC & 2021/04/23 & 187862 & 14.2 & 9391500002 & FWC & 2021/04/26 & 301412 & 13.9 \\
9392100002 & FWC & 2021/05/08 & 468470 & 13.8 & 9392200002 & FWC & 2021/05/09 & 941516 & 13.6 \\
9392400002 & FWC & 2021/05/13 & 855626 & 13.7 & 9392600002 & FWC & 2021/05/17 & 822558 & 13.7 \\
9392900002 & FWC & 2021/05/23 & 262773 & 14.4 & 9393000002 & FWC & 2021/05/25 & 581991 & 14.4 \\
9393100003 & FWC & 2021/05/27 & 285148 & 13.9 & 9393300002 & FWC & 2021/05/31 & 1198966 & 14.3 \\
9393800003 & FWC & 2021/06/11 & 871319 & 14.0 & 9394500003 & FWC & 2021/06/24 & 640088 & 14.3 \\
9395000003 & FWC & 2021/07/04 & 239706 & 14.4 & 9395900002 & FWC & 2021/07/22 & 353774 & 14.1 \\
9395900003 & FWC & 2021/07/22 & 1172248 & 14.0 & 9396900002 & FWC & 2021/08/10 & 991042 & 14.0 \\
9396900003 & FWC & 2021/08/11 & 987147 & 14.1 & 9397000002 & FWC & 2021/08/12 & 990990 & 13.9 \\
9397000003 & FWC & 2021/08/13 & 977884 & 14.0 & 9397100002 & FWC & 2021/08/14 & 332194 & 14.4 \\
9397100005 & FWC & 2021/08/15 & 259719 & 14.3 & 9397200002 & FWC & 2021/08/16 & 213972 & 14.1 \\
9397200005 & FWC & 2021/08/17 & 238072 & 14.2 & 9397300002 & FWC & 2021/08/18 & 285510 & 14.5 \\
9397300005 & FWC & 2021/08/19 & 768307 & 14.7 & 9399300003 & FWC & 2021/09/28 & 385239 & 13.4 \\
9400500003 & FWC & 2021/10/22 & 210041 & 13.8 & 9401500004 & FWC & 2021/11/11 & 413549 & 14.1 \\
9401900002 & FWC & 2021/11/19 & 953019 & 13.5 & 9402300003 & FWC & 2021/11/26 & 1012097 & 13.9 \\
9403300002 & FWC & 2021/12/16 & 368013 & 13.5 & 9403400003 & FWC & 2021/12/19 & 178165 & 13.5 \\
9403500002 & FWC & 2021/12/20 & 558016 & 13.8 & 9403600003 & FWC & 2021/12/22 & 411436 & 13.4 \\
9404500002 & FWC & 2022/01/09 & 558999 & 13.5 & 9404900002 & FWC & 2022/01/17 & 1169373 & 13.5 \\
9405400002 & FWC & 2022/01/27 & 343044 & 13.3 & 9405500002 & FWC & 2022/01/29 & 339128 & 13.3 \\
9405600002 & FWC & 2022/01/31 & 809146 & 13.7 & 9405700002 & FWC & 2022/02/02 & 997842 & 13.0 \\
9405800002 & FWC & 2022/02/04 & 298712 & 13.3 & 9405900002 & FWC & 2022/02/06 & 926728 & 13.2 \\
9406000002 & FWC & 2022/02/09 & 557540 & 13.6 & 9406600003 & FWC & 2022/02/20 & 691895 & 13.5 \\
9406700004 & FWC & 2022/02/23 & 439510 & 12.8 & 9406900002 & FWC & 2022/02/26 & 998593 & 13.2 \\
9408100003 & FWC & 2022/03/22 & 462685 & 13.3 & 9408200003 & FWC & 2022/03/24 & 484379 & 13.2 \\
9409500002 & FWC & 2022/04/19 & 907558 & 13.1 & 9409800003 & FWC & 2022/04/25 & 496551 & 12.4 \\
9410500004 & FWC & 2022/05/10 & 374369 & 11.9 & 9410800007 & FWC & 2022/05/15 & 382594 & 12.8 \\
9411400002 & FWC & 2022/05/28 & 236706 & 12.1 & 9414500002 & FWC & 2022/07/28 & 1204891 & 10.5 \\
9414600004 & FWC & 2022/07/30 & 173260 & 10.9 & 9416000002 & FWC & 2022/08/27 & 185959 & 10.2 \\
9417500002 & FWC & 2022/09/25 & 524943 & 10.3 & 9417600002 & FWC & 2022/09/27 & 776095 & 9.7 \\
9420200002 & FWC & 2022/11/18 & 328076 & 10.6 & 9420600005 & FWC & 2022/11/27 & 254437 & 10.2 \\
9423600005 & FWC & 2023/01/26 & 281947 & 9.5 & 9423900005 & FWC & 2023/02/01 & 794129 & 9.7 \\
9424200002 & FWC & 2023/02/06 & 499062 & 9.9 & 9424600003 & FWC & 2023/02/15 & 839094 & 9.4 \\
9427400002 & FWC & 2023/04/11 & 412498 & 9.1 & 9427800002 & FWC & 2023/04/19 & 175099 & 8.6 \\
9427900003 & FWC & 2023/04/22 & 893408 & 8.8 & 9430300003 & FWC & 2023/06/08 & 459575 & 8.5 \\
9431100005 & FWC & 2023/06/24 & 328207 & 8.5 & 9431400003 & FWC & 2023/06/30 & 243570 & 8.1 \\
9433900003 & FWC & 2023/08/19 & 404258 & 7.4 & 9435600002 & FWC & 2023/09/22 & 522256 & 7.0 \\
9436200002 & FWC & 2023/10/03 & 1164979 & 7.3 & 9437100002 & FWC & 2023/10/21 & 236664 & 7.0 \\
\end{longtable}

\end{spacing}
\end{document}